\documentclass{article}

\usepackage{amsmath,amssymb}
\usepackage{graphicx}
\DeclareGraphicsExtensions{.eps .pdf .jpg}
\usepackage[nolists,nomarkers]{endfloat}

\usepackage{hyperref}

\usepackage{fullpage}

\newcommand{\xx}[2]{y_{#1}^{(#2)}}

\newcommand{\vo}{{V_0}}

\newcommand{\wo}{{W_0}}

\newcommand{\xk}[1]{{\xx{k}{#1}}}

\newcommand{\kep}[1]{{\epsilon}^{(#1)}}
\newcommand{\fkep}[1]{\tilde{\epsilon}^{(#1)}}

\newcommand{\dt}{{\partial_t}}

\newcommand{\cD}[1]{\mathcal{D}_{#1}}
\newcommand{\rem}[1]{{}}
\newcommand{\mean}[1]{\left<{#1}\right>}
\newcommand{\spike}{\text{spike}}
\newcommand{\STA}{\text{STA}}

\newcommand{\cnd}[2]{\left({#1}| {#2}\right)}
\renewcommand{\ss}{{\bf s}}

\newcommand{\zz}{{\bf z}}
\newcommand{\jj}{{\bf j}}
\newcommand{\yy}{{\bf y}}

\begin{document}
\begin{center}
{\large \bf Single neuron computation: from dynamical system to
feature detector}

\bigskip\bigskip

{Sungho Hong$^1$, Blaise Ag\"uera~y~Arcas,$^2$ and Adrienne L.
Fairhall,$^{1}$

\medskip
$^1$Department of Physiology and Biophysics, \\
University of Washington, Seattle WA 98195 \\
$^2$Program in Applied and Computational Mathematics, \\
Princeton University, Princeton, New Jersey 08544 \\
\medskip
\{shhong,fairhall\}@u.washington.edu, blaisea@microsoft.com
\bigskip}\\

\end{center}

{\flushleft {\large \bf Abstract}} \\
\hspace{2cm} \\
White noise methods are a powerful tool for characterizing the
computation performed by neural systems. These methods allow one to
identify the feature or features that a neural system extracts from
a complex input, and to determine how these features are combined to
drive the system's spiking response. These methods have also been
applied to characterize the input/output relations of single neurons
driven by synaptic inputs, simulated by direct current injection. To
interpret the results of white noise analysis of single neurons, we
would like to understand how the obtained feature space of a single
neuron maps onto the biophysical properties of the membrane, in
particular the dynamics of ion channels.  Here, through analysis of
a simple dynamical model neuron, we draw explicit connections
between the output of a white noise analysis and the underlying
dynamical system. We find that under certain assumptions, the form
of the relevant features is well defined by the parameters of the
dynamical system. Further, we show that under some conditions, the
feature space is spanned by the spike-triggered average and its
successive order time derivatives.

\section{Introduction}
\label{sect:intro} A primary goal of sensory neurophysiology is to
understand how stimuli from the natural environment are encoded in
the spiking output of neurons. A useful tool for performing this
characterization is white noise analysis
\cite{marmarelis,wiener,spikes}, whereby a system is stimulated with
a randomly varying broadband signal and the relevant features
driving the system are determined by correlating output spikes with
the signal.  Extensions of this analysis to second order have been
used to find multiple stimulus features to which the system is
sensitive \cite{deruyter&bialek88,naama, bialek&deruyter05}.
Recently, these methods have been applied to characterize the
computation performed by a single neuron on its current inputs
\cite{hh,i&f,slee}.

Due to detailed knowledge of the dynamics of ion channels, it is
possible to build dynamical models of current flow in neurons that
accurately reproduce the voltage response of single neurons to
current (or conductance) inputs.  White noise analysis provides us
with the capacity to reduce this complex set of dynamical equations
to a simple functional model with concrete components: linear
feature selection followed by a nonlinear decision function
generating spikes. This procedure is compelling as it provides a
clear intuition about the form of the changes in the stimulus to
which the system is sensitive, and how the system combines these
features in the decision to fire. White noise methods have given
insight into coding properties at the systems level, in visual
\cite{naama,fairhallRetina,touryan,rust,horwitz} and somatosensory
\cite{rasAbstract,maravall} cortex. More recently, these methods
have also been applied to characterize neural coding in single
auditory \cite{slee}, central \cite{randy&marcARMA} and model
\cite{hh,i&f} neurons. However, mostly missing from these analyses
is a clear link between the obtained stimulus features and the
biophysical properties of the circuit or neuron. This issue is
particularly well-posed for single neurons under stimulation by
direct somatic current injection, where the biophysical parameters
of the soma must dominate the neuron's computation. Thus, given the
power of white noise methods, the questions that we wish to address
here are: how does the dynamical system governing neural behavior
map to the neuron's functional characterization derived using white
noise analysis? How are the features determined by the neuron's
biophysical properties?

\subsection{Minimal spiking models}\label{sec:minimal-spiking-models}

\begin{figure}[tp]\centering
\includegraphics[scale=1]{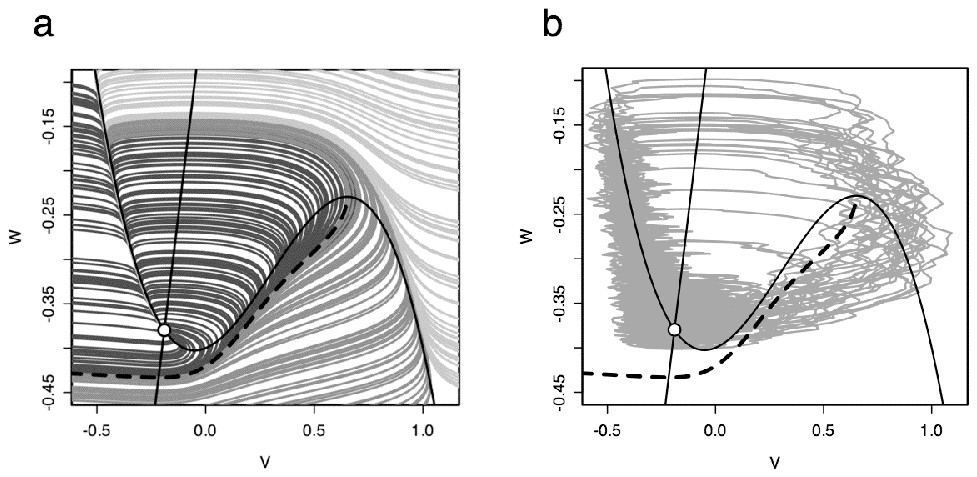}
\caption{(a) Phase plane of the FitzHugh-Nagumo neuron model with
trajectories and threshold (dashed). Parameters are $a = 0.1$,
$b=0.5$, $c=-0.4$ and $\psi=0.015$. (b) A single extended trajectory
on the FN phase plane driven by white noise input, showing multiple
spikes. The threshold (dashed) still makes an approximate boundary of
the spiking and non-spiking portions of the trajectory (gray).}
\label{fig:FNphaseplane}
\end{figure}

Neural dynamics are typically described by conductance-based models
which describe the temporal evolution of the voltage $V$ due to an
input current $I$ and the variable ionic conductances of ion
channels:
\begin{equation}\label{eq:ne1}
C dV/dt = -\sum_i g_i (V) (V - E_i) + I,
\end{equation}
where $C$ is the membrane capacitance.  Each conductance type $i$
has a reversal potential $E_i$, where the conductances $g_i$
corresponding to different ion channels may be voltage dependent
through the dynamics of the corresponding activation and
inactivation gating variables $\xi_i$ and $\phi_i$:
\begin{eqnarray}
g_i & = & \bar{g}_i \xi_i^{p_i} \phi_i^{q_i}, \label{eq:ne2}\\
d\xi_i/dt & = & f_{\xi_i}(\xi,V)\\
d\phi_i/dt & = & f_{\phi_i}(\phi,V),
\end{eqnarray}
where $\bar{g}_i$ are constants, and the functions $f_\xi$ and
$f_\phi$ are affine in $\xi$ and $\phi$ respectively, but not in
$V$. The exponents $p_i$ and $q_i$ are usually taken to be integers,
and describe the cooperativity of molecules required to gate the ion
channel.  This set of equations is highly nonlinear due to this
cooperativity and the voltage dependence of the conductances. While
ultimately we would like to understand systems of many dynamical
variables, we will begin by analyzing a simplified spiking system,
the FitzHugh-Nagumo (FN) neural model \cite{fitzhugh,nagumo}. The
FitzHugh-Nagumo system is cubic in $V$, the lowest order
nonlinearity supporting spike-like behavior, and is defined by:
\begin{eqnarray}
    \psi \, dV/dt & = & V(1-V)(a+V) - W + c + I, \label{eq:fhna}\\
    dW/dt & = & V - bW, \label{eq:fhnb}
\end{eqnarray}
where $V$ is voltage, $W$ is a generic inactivation variable with
linear recovery dynamics, $I$ is the input current and $a$, $b$, $c$
and $\psi$ are parameters, $\psi \ll 1$. The nullclines of the
system are the curves along which $dV/dt=0$ (the V nullcline, a
cubic) and $dW/dt=0$ (the W nullcline, a straight line).
Fig~\ref{fig:FNphaseplane}a shows the nullclines for zero input $I$
on the phase plane.  Trajectories have been traced forward and
backward in time through a randomly chosen set of initial
conditions. The unique intersection of the nullclines is the stable
fixed point. Elsewhere in the plane, trajectories always spiral
counterclockwise, but qualitatively there are two kinds of orbits,
ultimately both converging to the stable fixed point. Subthreshold
orbits make only small excursions; however, for trajectories
starting either with low inactivation (roughly $W<-.45$) or high
voltage (for example, $V>0$), the system makes a large excursion
around the right branch of the $V$ nullcline, corresponding to a
spike, before settling back to the fixed point. Although we will
generally be considering a time-varying input $I(t)$, we will make
heavy use of the $I=0$ phase plane: one can regard this as
representing the instantaneous flow if the input current is switched
off.  For $I = 0$, one can define a separation of spiking initial
conditions from subthreshold conditions, marked in light vs. dark gray in
Fig~\ref{fig:FNphaseplane}a. This separation of behavior can be
thought of as a {\em threshold}. Typically the threshold for spiking
is taken to be a threshold in voltage only, but this phase plane
picture demonstrates that one should really consider the threshold
to depend on both $V$ and the hidden inactivation variable(s).

One can approximate this ``dynamical'' threshold by considering a
minimal spiking trajectory.  There is no absolute distinction
between spiking and subspiking trajectories in type II neurons
\cite{gerstnerSRM} such as Hodgkin-Huxley and FitzHugh-Nagumo, so
this choice is necessarily somewhat arbitrary.  Starting with a
point on the phase plane at the end of the spike plateau, one can
eliminate time from the dynamical equations, solving parametrically
for the curve passing through this point. Here, we obtained one such
curve by numerically solving the equations from the local maximum of
the cubic nullcline, given by $\frac{d}{dV}V(1-V)(a+V) = 0$.

One of the issues we will explore in this paper is the relationship
between the geometry of the threshold in the multidimensional space
of the dynamical variables and the measurements one makes in white
noise analysis.

\subsection{White noise analysis}

Our approach will be based on the functional perturbation expansion
method, which we will briefly review here. The simplest such
expansion is known as the Volterra series
\cite{volterra,marmarelis,marmarelisBook}, where the functional
$y(t) = y[I(t)]$ is expanded as
\begin{equation}
y(t) = h_0 + \int dt_1\, h_1(t_1) I(t-t_1) + \frac{1}{2!}\int dt_1
dt_2\, h_2(t_1,t_2) I(t-t_1) I(t-t_2) + \cdots,
\end{equation}
where the kernels $h_i$ are called Volterra kernels. One problem
with Volterra analysis is that the successive order kernels are not
independent. For example, $h_2$ contains a constant component $\sim
\int dt_1 \, h_2(t_1,t_1)$.

Wiener analysis \cite{wiener,spikes} avoids this issue. In this
framework, $I(t)$ is assumed to be zero mean Gaussian white noise
satisfying
\[
\left< I(t) I(t') \right > = \sigma^2 \delta(t-t').
\]
Under Wiener analysis, $y[I(t)]$ is expanded as
\begin{eqnarray}
y(t) &=& g_0 + \int dt_1\, g_1(t_1) I(t-t_1) \nonumber\\
&& \quad + \frac1{2!}\left\{ \int dt_1 dt_2\, g_2(t_1,t_2) I(t-t_1)
I(t-t_2) - \sigma^2 \int dt'\, g_2(t',t')\right\}  +
\cdots.\label{eq:volterra} \end{eqnarray} In the Wiener expansion,
each kernel is independent from the others, since at each order, one
subtracts out the lower order components. Therefore, the kernels
$g_i$ can be easily obtained by measuring the correlation functions
between $y(t)$ and $I(t)$. For example,
\begin{eqnarray*}
\mean{y(t)} &=& g_0 + \int dt_1\, g_1(t_1) \mean{I(t-t_1)} \\
&& \quad +\frac{1}{2!}\left\{ \int dt_1 dt_2\, g_2(t_1,t_2) \mean{I(t-t_1) I(t-t_2)} -
\sigma^2 \int dt'\, g_2(t',t')\right\}  + \cdots\\
& = & g_0,\\
\mean{y(t) I(t-t_1)} & = & g_1(t_1)/\sigma^2, \\
\mean{y(t) I(t-t_1) I(t-t_2)} & = & g_2(t_1,t_2)/\sigma^4,
\end{eqnarray*}
and so on. The correlation method can be used to determine the relationship
between Wiener and Volterra kernels. From Eq~\eqref{eq:volterra},
\begin{eqnarray*}
g_0 & = & h_0 + \int dt_1\, h_1(t) \mean{I(t-t_1)}\\
&&\quad+\frac{1}{2!}\int dt_1 dt_2\, h_2(t_1,t_2) \mean{I(t-t_1) I(t-t_2)} +
\cdots\\
& = & h_0 + \frac{\sigma^2}{2!} \int dt' \,h_2(t',t') + \cdots,\\
g_1(t) & = & h_1(t) + \frac{\sigma^2}{2!}\int dt' h_3(t',t',t) + \cdots,
\end{eqnarray*}
and so on.

Wiener analysis has been successfully used to determine the relevant
feature space for neural systems by reverse correlation of the
spiking output with white noise input. For this application, the
output is taken to be the sequence of spike times, $\rho(t) = \sum_i
\delta(t-t_i)$.  Let us define the stimulus ${\bf s}$ as a linear
transformation of the current input $I(t)$:
\begin{equation}
s_i = \int dt'\, f_i(t') I(t-t') \label{eq:sdef}
\end{equation}
where $f_i$ are some set of independent linear filters. The goal is
to determine how a single spike is generated by  $I$, reduced to a
vector $\bf s$ comprising projections along the {\em relevant}
dimensions $f_i$.

We need to compute the probability of spiking as a function of the
input, $P\cnd{\spike}{{\bf s}}$:
\begin{equation}
P\cnd{\spike}{{\bf s}} = P(\text{spike}) G({\bf s}),
\end{equation}
where $G({\bf s})$ is the input/output function relating $\rho(t)$
to the time-varying stimulus ${\bf s}(t)$, and we have separated out
the scale factor $P(\text{spike})$ proportional to the mean firing
rate. In principle, since the equations of motion are deterministic,
$P\cnd{\spike}{{\bf s}}$ is a Boolean function. However, as we will
be approximating ${\bf s}$ by a finite dimensional truncation, and
will be solving for $P$ itself in successive orders, the result will
no longer be Boolean.

We can compute $P\cnd{\spike}{{\bf s}}$ using Bayes' theorem
\cite{spikes},
\begin{equation}\label{eq:defgs}
G({\bf s}) = \frac{P\cnd{\spike}{{\bf s}}}{P(\spike)} =
\frac{P\cnd{{\bf s}}{\spike}}{P({\bf s})}.
\end{equation}
This shows that the Wiener kernels of $G({\bf s})$ are the moments
of $P\cnd{{\bf s}}{\spike}$, which can be easily measured. In this
paper, we will focus on the first two moments\footnote{Note that the
zeroth order kernel is simply $g_0=1$.} of this distribution. The
first moment is the spike-triggered average (STA), $\bar{s}(t)$:
\begin{equation}
\bar{s}(t) = \left<I(t_i - t)\right>_i = \int\! d{\bf s}\, \, {\bf
s} P\cnd{{\bf s}}{\spike} =  \int \! d {\bf s} \,\, {\bf s} P({{\bf
s}}) G({\bf s}),
\end{equation}
where the average $\left<\cdot\right>_i$ over the spike occurrence
times $\{t_i\}$ is replaced with an average over the
spike-triggering ensemble. When the neuron has multiple
characteristic features, the STA may contain little information
about them since it only points in a single direction which is a
certain linear combination of relevant features. The second order
kernel, however, can give such information. This kernel is related
to the spike-triggered covariance (STC) matrix $C$, which we will
define as
\begin{eqnarray} C(t,t') &=&
\mean{I(t_i-t)-\bar{s}(t),I(t_i-t')-\bar{s}(t')}_i
- \mean{I(\tau-t)I(\tau-t')}_\tau \nonumber\\
& = & \sigma^2 g_0\delta(t,t') + \sigma^4 g_2(t,t') -
\bar{s}(t)\bar{s}(t') -\sigma^2\delta(t,t')\\ \nonumber  & = &
\sigma^4 g_2(t,t') - \bar{s}(t)\bar{s}(t'). \label{eq:covarianc}
\end{eqnarray}
$g_2(t,t')$ is the second order Wiener kernel of $G({\bf s})$.  When
a functional $F[{\bf s}]$ depends on Gaussian noise ${\bf s}$,
$\mean{s_i F[{\bf s}]} = \sum_j \mean{s_i s_j}\mean{\partial_j
F[{\bf s}]}$. Thus, we can expand $g_2$ \cite{bialek&deruyter05}:
\begin{equation} g_2(t,t') =
\sum_{i,j} \mean{\frac{\partial^2 G}{\partial s_i
\partial s_j}} f_i (t) f_j(t'), \label{eq:G2}
\end{equation}
where Eq.~\eqref{eq:sdef} introduces the linear filters which define
the components of ${\bf s}$.

Therefore, unless the modified Hessian $\hat{H}_{ij} =
\mean{\frac{\partial^2}{\partial s_i \partial s_j} G}-
\mean{\frac{\partial}{\partial s_i} G}
\mean{\frac{\partial}{\partial s_j} G}$ has null
directions\footnote{Since $\bar{s}(t)\bar{s}(t')$ is only rank one,
there are only two possibilities leading to null directions. The
first is when $\mean{\partial_i \partial_j G}$ has null directions,
in which case it is excluded (for our purposes) by definition. The
second case is when $\mean{\partial_i \partial_j G}$ has an
eigenvector $\bar{s}(t)$ with an eigenvalue $|| \bar{s}(t) ||^2$,
which does not happen generically.}, the STC is spanned by the
relevant feature space $\{ f_i \}$. Diagonalizing the STC and
extracting the eigenvectors with largest absolute eigenvalue will
determine the leading dimensions spanning the feature space
\cite{naama,hh,bialek&deruyter05}. Eigenmodes appearing with
negative eigenvalue correspond to stimulus directions in which the
variance of the spike-conditional distribution is reduced relative
to the stimulus prior distribution; eigenmodes with positive
eigenvalues are directions with increased variance.  Such cases may
arise if the spike-conditional distribution is bimodal in some
direction, or if the spike-conditional distribution forms a ring
around the origin; phase-invariant (complex) cells are an example
\cite{simoncelli,rust,fairhallRetina,rasAbstract,maravall}.

In this paper, we work with a multidimensional space of dynamical
variables, as in Eq.~(1-6), for which we obtain the Volterra series.
When we denote them with $y_i$, the Wiener kernels for $G(y_i)$ can
be expanded as
\begin{eqnarray}
g_1[G] &=& \sum_i \frac{\partial G}{\partial y_i} g_1[y_i] + \cdots,
\label{eq:1stwieneriny} \cr
g_2[G] & = & \sum_i {\partial G \over \partial y_i} g_2[y_i] +
\sum_{i, j}\frac{\partial^2 G}{
\partial y_i \partial y_j} g_1[y_i] g_1[y_j] + \cdots\label{eq:2ndwieneriny}
\cr &\cdots\nonumber&
\end{eqnarray}
where $g_n[y_i]$ is a $n$th Wiener kernel of $y_i$ and dots
represent higher order terms than quadratic. Here we note that
$g_n[y_i]$ is generated by the infinite series of Volterra kernels.
Moreover, we will see later that the Volterra kernels at each order
can be constructed from the set of first order kernels.

\section{Linear systems}
\label{sect:linear}

In the previous section, we discussed the application of white noise
methods to experimental data where the output of the system is
measured as the firing times of spikes. With access to the entire
dynamical system, one could attempt to model the relationship
between the input $I(t)$ and the output voltage $V(t)$, including
both subthreshold and spiking behavior. As we have discussed,
neurons are highly nonlinear. Perturbative expansions such as
Wiener/Volterra series generally are unable to capture the
nonlinearities of spike generation with few terms. However, a very
successful and highly influential approach to the modeling of
neural systems has been to separate a linear filtering stage from an
explicit nonlinearity capturing the very sharp voltage increase of
the spike \cite{victor&shapley,gerstnerSRM,gerstner,berry&meister}.
This is known as a cascade model, and in various forms is the
theoretical basis for many of the simple neuron models in use,
including the integrate-and-fire neuron \cite{gerstnerSRM,i&f}, the Spike
Response Model \cite{gerstnerSRM}, generalized integrate-and-fire
models \cite{paninski&pillow} and cascade models of retinal ganglion
cells \cite{victor&shapley,victor&shapleyLNL,keat,pillowRetina}.
The success of these models suggests that a breakdown of the
dynamical system into its linear component and a threshold is a
useful approximation to examine. In our analysis of the dynamical
system, we will take the approach of experimental work, where the
output is reduced to a sequence of spike times.  Our goal here is to
derive the form of the multidimensional linear stage that arises
directly from the equations of motion, and to examine the
consequences of thresholds of various forms.

We will begin by walking through some simple linearized systems
analysis and make clear the connection with the cascade model
picture.  Rewriting Eqs. (1-4) in the following simple form,
\begin{equation} \label{eq:lsystem}
\dt y_k = f_k ( y_1, y_2, \ldots, y_N ) + I(t)\delta_{k0}, \quad k = 0, 1,
\cdots, n-1,
\end{equation}
we assume that they have a fixed-point solution $y_k ( t ) =
y_k^{(0)}$ when $I(t)=0$. Now the linear approximation of the system
around the fixed point is given by\footnote{From here on, we use the Einstein
summation convention, e.~g.~$C_{km}d_m \equiv \sum_m C_{km} d_m.$}
\begin{equation}\label{eq:7}
\cD{km}\xx{m}{1} =   I(t)\delta_{k0},
\end{equation}
where
\begin{equation}\label{eq:3}
\quad \cD{km} = \delta_{km}\dt - J_{km},\quad J_{km} = \left.
\frac{\partial f_k }{\partial y_m}\right|_{y^{( 0 )}}.
\end{equation}
The linear filters are the kernels of Eq~\eqref{eq:3}, $\kep{1} =
\cD{}^{-1}$. They satisfy
\begin{equation}\label{eq:9}
\cD{km}\kep{1}_{ml}(t) = \delta_{kl}\delta ( t ),
\end{equation}
giving the linear filter form for the system's evolution:
\begin{equation}
\label{eq:filts} \xk{1} ( t ) = \int^t_{-\infty} dt' \, \kep{1}_{k}
( t - t' ) I ( t' ).
\end{equation}
where $\kep{1}_{k} = \kep{1}_{k0}$.
Eq~\eqref{eq:9} can be solved by diagonalizing the Jacobian matrix
$J_{km}$. Let's consider the following diagonalization with
eigenvalues $\lambda_\alpha$,
\begin{equation}\label{eq:diagonalize}
\lambda_{\alpha}\delta_{\alpha \beta} = U_{\alpha k} J_{km} U^{-1}_{m \beta}.
\end{equation}
Here we employ the notation that indices in the diagonalized basis
are Greek letters. Therefore, Eq~(\ref{eq:9}) can be written in the
new basis as
\begin{equation}\label{eq:11}
\cD{\alpha} \fkep{1}_\alpha = U_{\alpha 0}\delta(t), \quad
\cD{\alpha} = \dt - \lambda_\alpha, 
\end{equation}
where the first order kernels $\fkep{1}$ in this basis are
\[
\fkep{1}_\alpha(t)= e^{\lambda_\alpha t}H(t),
\]
where $H(t)$ is the Heaviside step function and arises from
causality. The eigenvalues $\lambda_\alpha$ can be real or complex.
The sign of the real part of the eigenvalues is determined by the
stability properties of the fixed point at $y^{(0)}$. For a neural
system this fixed point should be attracting, so all eigenvalues
have negative real part and thus eigenvectors decay as $t\rightarrow
\infty$. This underscores that there are only a limited number of
forms that the linear filters can take: decaying exponentials or
damped oscillations.

We note the following facts. First of all, in general, an
$n$-dimensional system will have $n$ independent first order
kernels, except when the Jacobian has a degenerate spectrum.
Further, the linear kernels can be generated from a ``master
kernel''
\[ \varphi(t) = \sum_\alpha e^{\lambda_\alpha t} H(t)
\label{eq:masterfilt}
\]
and its time derivatives up to $n$th order. To show this, from the
expression of the derivative of a master kernel
\[
\dt^k \varphi(t) = \sum_\alpha \lambda_\alpha^k e^{\lambda_\alpha t}
=  T_{k \alpha} e^{\lambda _\alpha t}, \qquad T_{k \alpha} =
\lambda_\alpha^k,\ (k, \alpha = 0,\ldots n-1),
\]
up to additive singular terms coming from the derivatives of $H(t)$.
$T_{k \alpha}$ is a Wronskian matrix of exponentials at $t=0$. The
determinant of $T_{k\alpha}$ is the Vandermonde determinant
$\det(T_{k \alpha}) = \prod_{\alpha>\beta} (\lambda_\alpha -
\lambda_\beta)$, which cannot vanish by
definition~\cite{GradshteynRyzhik}. This means that the master
kernel and its derivatives can be transformed to $e^{\lambda_\alpha
t}$'s in a non-singular way.

Therefore, we conclude that all first order kernels can be expressed
in terms of a master kernel and its time derivatives. Frequently the
first order kernel is non-zero at $t = 0$, so that the time
derivative of the filter also includes a delta function at $t = 0$.
We will see the appearance of this singular component in the white
noise analysis.

To recover the filters corresponding to the dynamical variables, one
must invert the diagonalizing operation, Eq.~\eqref{eq:diagonalize}
to obtain linear combinations of the eigenmodes:
\begin{equation}\label{eq:k1final}
\kep{1}_i(t) = U^{-1}_{i\alpha} \fkep{1}_\alpha(t) =  \sum_\alpha
c_\alpha e^{\lambda_\alpha t}H(t),
\end{equation}
where the coefficients $c_\alpha$ derive from the components of the
Jacobian. Thus the filters of the linear system are sums of
exponentials, either purely real or with an imaginary component. To
the extent that subthreshold dynamics are well approximated by the
linearized system, this shows clearly why one might expect to find
both integrate-and-fire-like neurons, with filters having purely
real associated eigenvalues, and oscillate-and-fire neurons
\cite{oscillate-and-fire,hutcheonYarom} with a corresponding
eigenvalue having nonzero imaginary part. In terms of the linear
system, the possible set of filters or feature space of the system
is dual with the dynamical variables which define the system. The
subset of features that are relevant to a spike occurring are those
that contribute to crossing the internal state over ``threshold''.

The properties outlined above have an interesting implication for
white noise analysis in the case that the system is well described
by subthreshold linearity. As we will see in the next section, the
STA is typically a linear combination over the relevant features and
their derivatives. If the STA has components onto all the true
features, as it typically will, one can derive the basis for the
feature space from the STA and its successive order time
derivatives, as has been empirically observed \cite{fredthesis}.

\subsection{Higher order series terms}

The higher order approximations to the system can also be expressed
in terms of first order kernels. For example, the second order
approximation is given by the following equation:
\begin{equation}\label{eq:8}
\cD{km} \xx{m}{2} =  \frac{1}{2!}\left. H_{kmn} \xx{m}{1}
\xx{n}{1},\quad H_{kmn} = \frac{\partial^2 f_k}{\partial y_m
\partial y_n}\right|_{y^{( 0 )}},
\end{equation}
whose solution is
\begin{equation*}
\xk{2} = \frac{1}{2!}
\int ds_1 ds_2 \, \kep{2}_{k}( t- s_1,t- s_2 ) I ( s_1 ) I ( s_2 ),
\end{equation*}
\begin{equation}\label{eq:2ndkernel}
\kep{2}_{k} ( t_1, t_2 ) =
\int dt'\,H_{lmn}
\kep{1}_{kl} ( t' )
\kep{1}_{m} ( t_1-t' )
\kep{1}_{n} ( t_2-t' ).
\end{equation}
In this way, all of the higher order kernels can be obtained by
outer products of first order kernels. The detailed computation is
summarized in appendix~\ref{sec:volt-expans-dynam}.

\subsection{White noise analysis of threshold-crossing linear neurons}
\label{sec:white-noise-analysis} In this section, we will assume
that the higher order nonlinearity of the system is well captured by
a threshold. Since voltage is generally the only observed variable,
``the threshold'' is often taken to be a threshold on the voltage.
The imposition of a threshold on the voltage alone has been applied
even to neural models such as FN and the Hodgkin-Huxley model
system, where there is no difficulty in defining a threshold in $V$
and the ``hidden" dynamical variables of $W$ (FN) or $m$, $n$, and
$h$ (HH). Here we will consider the implications of considering the
threshold to apply in multiple dimensions. Before we discuss the
validity of the threshold approximation for a FN neuron, we will
illustrate the results from applying white noise analysis to models
with a variety of threshold structures acting on subthreshold linear
dynamics given by the linearized FN system.

Previous work has treated simple one-dimensional linear/nonlinear
models, in which a spike is defined when the output of a single
linear filter $f$ on a random current input crosses a threshold. One
can show \cite{deWeeseThesis}[Meister, M., private communication]
that covariance analysis on
this model finds two modes, the filter $f$ and its time derivative,
$f'$. The derivative mode is a result of the criterion defining the
spike only on the upward crossing of the threshold:
\begin{equation}
\frac{d}{dt} \int^t d\tau f(t-\tau) I(\tau) > 0 \quad \rightarrow
\quad \int^t d\tau f'(t-\tau) I(\tau) > 0.
\end{equation}
Thus the variance of projections onto $f'$ is also reduced with
respect to the prior. Covariance analysis has also been performed
\cite{fairhallRetina} on more realistic models in which an
exponentially decaying afterhyperpolarization is added to the
voltage following a spike \cite{gerstnerSRM,keat}.  This makes the
threshold-crossing model more realistic, but the manipulation does
not alter the $(f,f')$ eigenmode structure unless the timescale of
the AHP is short enough that spiking occurs many times during the
superthreshold fluctuation, which destroys the correlation with
positive $f'$.

Here we will assume that the system evolves according to linear
subthreshold dynamics. We treat five cases. In the first three, the
threshold is linear, but in different components of the 2D space; in
the last two, the threshold is a nontrivial 2-dimensional structure.
The thresholds used are:
\begin{enumerate}
\item a traditional threshold in $V$ only
\item a threshold in $W$ only
\item a threshold on a linear combination of $V$ and $W$
\item the union of two piecewise linear segments in $V$ and
$W$, i.e. a logical `or' on 1 and 2
\item a curved threshold on a smooth, nonlinear function of $V$ and $W$.
\end{enumerate}
From the discussion in \S\ref{sect:linear}, the linear space in
which the $n$-dimensional system operates is defined by the $n$
filter inversions of the dynamical system, Eq.~\eqref{eq:filts}. The
effect of the threshold is to select as {\em relevant} from those
$n$ dimensions the subspace defined by the filter directions that
have a nonzero projection onto the threshold. Thus we would predict
that for the linear threshold, we will find two relevant modes, the
primary direction $f$ which is normal to the direction of the
threshold, and $f'$, its time derivative. For a threshold in a
single dimension, $V$ or $W$ only, $f$ should recover the $V$ or $W$
filter respectively, and the corresponding time derivative. For a
threshold that is a linear combination of $V$ and $W$, $f$ should be
a linear combination of the $V$ and $W$ filters. The distribution of
trajectories is simply a linear transformation of the original
Gaussian stimulus distribution. Any linear threshold in this plane
will produce negative eigenmodes, as the set of points selected by
the threshold must have decreased variance with respect to this
prior.

When the threshold is not linear in $V$ and $W$, we expect to find
two primary filters $f_1$ and $f_2$ which will be some linear
combination of the $V$ and $W$ filters, and in principle, the time
derivatives. However, the number of modes will be less than or equal
to $n+1$ because, as we have shown, the time derivatives span the
same space as the linear variables themselves. (Note that the $+1$
accounts for the singular component of the time derivative,
$\delta(t)$, arising from discontinuity of the filters at $t=0$.)
Furthermore, a threshold which is not linear may produce a
spike-conditional distribution with a direction in which the
variance is actually increased with respect to the prior; such a
direction will appear with a positive eigenvalue.

\begin{figure}[p]\centering
    \includegraphics[scale=1]{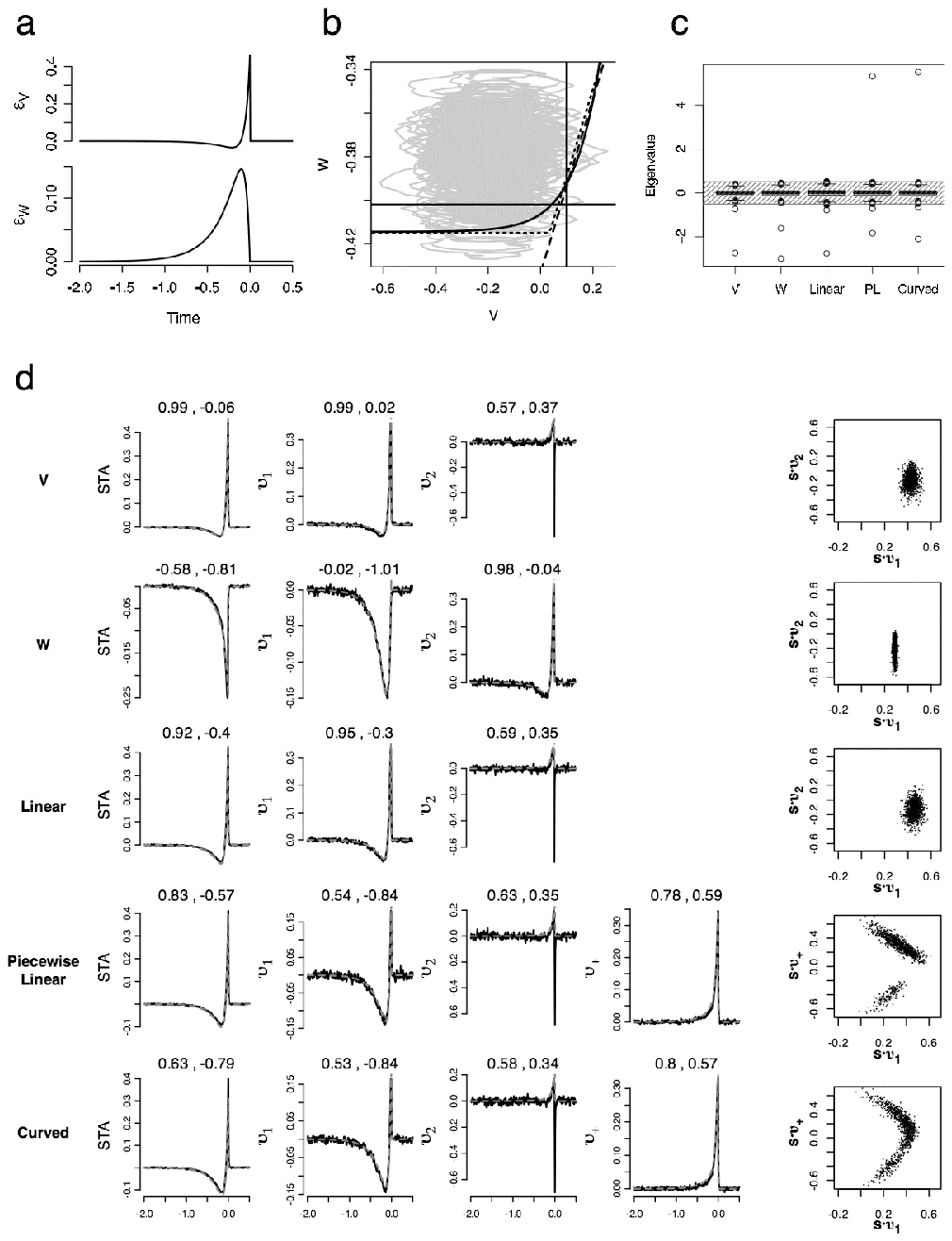}
\caption{ (a) First order kernels of the FN neuron model, drawn as
filters, i.~e. in $-t$.(b) Various thresholds for the linear system.
$V$ (vertical), $W$ (horizontal), linear (dashed), piecewise linear (dotted),
and curved (solid curve) thresholds are used. (c) Spectra of covariance
matrices for the thresholds. The shaded box represents the level of
error from finite sample size and dimension. (d) STAs and covariance
modes of the linear system with various thresholds. The non-singular
part of each mode is fitted using least squares to a linear
combination of the (orthogonalized) first order kernels as $v_{fit}
= c_V\kep{1}_{V}+c_W\kep{1}_{W}$. The kernels are normalized
excluding the $\delta$-function component. The coefficients $(c_V,
c_W)$ are displayed above each plot and the gray line is a fitted
function. For each case we show the projection of spike-triggered
current histories onto either the leading two negative modes, $v_1$
and $v_2$, or $v_1$ and $v_+$, depending on the threshold.
}
\label{fig:FNfirst}
\end{figure}

Fig~\ref{fig:FNfirst}b shows the thresholds that we applied
to the FN first order linear dynamics. The results of the covariance
analysis are shown in
Figs~\ref{fig:FNfirst}c and d.
Fig~\ref{fig:FNfirst}c shows the covariance eigenvalue spectra for
all five cases. The significant eigenvalues are empty circles beyond
the error level denoted by the shaded box.\footnote{Significance of
eigenvalues can be evaluated as follows. Let $N$ be the number of
spikes and $d$ the dimension of each sampled stimulus. To estimate
the finite size/dimension error, we choose $N$ $d$-dimensional
stimuli at random~\cite{rust}. In this case, the covariance matrix
$C$ in Eq.~\eqref{eq:covarianc} is filled with $d(d+1)/2$ random
numbers drawn from a normal distribution~${\cal
N}(0,\sigma^2\sqrt{2/N})$ in the large $N$ limit. When $d\gg1$, the
eigenvalues of this matrix follow Wigner's semicircle law, and their
distribution is bounded by $\pm 2\sqrt{2}\sigma^2\sqrt{d/N}$, which
is the error level~\cite{Mehta}. When the stimulus has small
correlation, the semicircle is not a good approximation of the
eigenvalue distribution, but change in the upper and lower bounds
rarely exceeds an order of magnitude, which can be checked by
numerical simulation.} In the cases with thresholds on
 $V$ only, on $W$ only, and linear in both $V$ and $W$,
only two modes are obtained. With 2D thresholds, cases 4 and 5,
three eigenmodes appear. Fig~\ref{fig:FNfirst}d shows the
corresponding modes. We see that the $V$ threshold picks out
$\kep{1}_V$, the $W$ threshold selects $\kep{1}_W$, and the
$(V,W)$-linear threshold leads to a linear combination of these
filters. In the latter two cases, the eigenmodes are a linear
combination of $\kep{1}_V$, $\kep{1}_W$ and $\dt \kep{1}_V$, itself
a linear combination of the $V$ and $W$ kernels and $\delta(t)$.

Fig~\ref{fig:FNfirst}d also shows the projection of spike triggered
stimuli onto two of the distinguished eigenvectors. In the first
three cases, the linearity of the threshold is manifested in the 2D
plane as a single elliptic cluster, while the non-linear threshold
cases show richer structure.

Recall that the filters are generated by linear combinations of a
master filter and its time derivatives. This implies that our
multidimensional threshold structure has a natural interpretation as
a {\em dynamical} threshold, depending not only on the voltage $V$, but
also on $\dt V$, $\dt^2 V$, and so on~\cite{azouz}. We will show how
well this approximation fits for the Hodgkin-Huxley model in the
final section.

\subsection{Analysis of threshold-crossing models in multiple dimensions}
An advantage of the simple models introduced in the previous section
is that they can be treated analytically. From Eq.~\eqref{eq:defgs},
we have
\[ G (\ss) =
\frac{P\cnd{\spike}{\ss} P(\ss)}{P ( \spike )} =
\frac{P\cnd{\spike}{\ss}
   P (\ss)}{\int D\ss\, P \cnd{\spike}{\bf{s}}  P
   (\ss)}, \]
where $\int D\ss= \int d s_1 d s_2 \cdots$. Instead of directly computing
$G(\ss)$, we characterize it by a moment generating function,
\begin{equation}\label{eq:defmomgen}
W [\jj] = \log \int D\ss P ( \spike
   |\ss) P (\ss) e^{\jj \cdot \ss} .
\end{equation}

We define an $n$-dimensional first order system, given by dynamical
variables $y_k$ and the corresponding linear filters
$\epsilon_k$ as in Eq.~\eqref{eq:7}. We assume that the stimulus is a Gaussian white
noise current $I(t)$ with zero mean and variance $\sigma^2$. Thus as
before,
\[
    y_k(t) = \int_0^{\infty} d \tau \epsilon_k ( \tau ) I ( t - \tau ) .
\]

We denote a random segment of the current stimulus $I(\tau), \tau
\le t$ as an infinite-dimensional\footnote{In practice, any
application to data requires discretization in time. We use the
convention that a function $f(t)$ is discretized as $\hat{f}_t =
f(t)\sqrt{\Delta t}$ where $\Delta t$ is the time step. This implies
$\int f^2 dt \approx \sum_i f(i\Delta t)^2 \Delta t = \sum_i
\hat{f}_i^2 = \hat{\bf f}\cdot \hat{\bf f}$ and thus the vector
$\hat{\bf f}$ obtained by discretizing $f(t)$ has a vector norm
corresponding to the $L^2$ norm.} vector ${\bf s}$. Any sample of
$\yy$ is therefore a functional of the random variable ${\bf s}$,
$\yy[{\bf s}]$; for simplicity we will simply write $\yy$.

Spiking is determined by crossing a threshold $\theta ( \yy ) = 0$
in the phase space from below. Then,
\[
    P\cnd{\spike}{{\bf s}} = \delta (\theta ( \yy ) ) H ( \dot{\yy} \cdot
    \nabla_\yy \theta ( \yy ) ) \dot{\yy} \cdot \nabla_\yy \theta ( \yy ),
\]
The Heaviside function $H(\cdot)$ ensures that spiking only occurs
on a threshold crossing from below, and the weight factor $\dot{\yy}
\cdot \nabla_\yy \theta ( \yy )$ is a geometric factor accounting
for the flux at the threshold. The filters $\epsilon_k$ are not
necessarily normalized or orthogonal to each other, and it is
convenient to define the orthonormal basis $\{ f_\mu \}$ which spans
the same space as $\{ \epsilon_k \}$. Then there will be a linear
transformation $T_{\mu k}$
\[
    f_{\mu} = T_{\mu k} \epsilon_k.
\]
and so we define a new coordinate system
\[
z_{\mu} \equiv T_{\mu k} y_k.
\]
As ${\bf s}$ is uniform Gaussian, the orthonormally transformed
variable $\zz$ is also uniform Gaussian with variance $\sigma^2$.

Now we can separate the stimulus into two components: its projection
into the subspace spanned by the $\{f_{\mu}\}$ and the orthogonal
component. We will denote the corresponding directions of $\jj$ as
$\jj = \jj_\| + \jj_\perp$, where
\[
j_{\| \mu}(t) = \int_0^\infty d\tau\, j ( t -\tau) f_{\mu} ( \tau).
\]
The moment generating function can then be separated as $W [ \jj ] =
W_{\|} [ \jj_{\|} ] + W_\perp [ \jj_\perp ]$ where
\begin{equation}
W_\perp [ \jj_\perp ] = \log \int D\ss_\perp\, e^{-s_\perp^2
 /2 \sigma^2 + \jj_\perp \cdot \ss_\perp} = \frac{\sigma^2}{2}j_\perp(t)^2+
\text{const},
   \label{eq:static2}
\end{equation}
where $\ss_\perp$ are the components of $\ss$ orthogonal to the
plane spanned by $\zz$, and
\begin{eqnarray}
W_\| [ \jj_\| ] &=& \log \int d^n \zz\, \det(T^{-1}) e^{- z^2 /2 \sigma^2}
\delta ( \theta ( \zz )
   ) H ( \dot{\zz} \cdot \nabla_\zz \theta ( \zz ) ) \dot{\zz} \cdot \nabla_\zz \theta (\zz ) e^{{\bf j}_\| \cdot \zz} \nonumber\\
& = & \log \int d^n \zz\, e^{-z^2/2\sigma^2} \delta ( \theta ( \zz )
) w ( \zz ) e^{{\bf j}_\|\cdot \zz} + \text{const}, \label{eq:static1}
\end{eqnarray}
with
\[ w ( \zz ) =
H ( ( R \zz ) \cdot \nabla_\zz \theta ( \zz ) ) ( R \zz ) \cdot \nabla_\zz
   \theta ( \zz ), \]
where $R = T J T^{-1}$ is the Jacobian matrix of the linear system,
Eq.~\eqref{eq:7}, in the $\{f_\mu\}$ basis, and we use
\[
 \dot{z}_\mu = T_{\mu k } \dot{y}_k = T_{\mu k} J_{k l} y_l =
 (TJT^{-1})_{\mu \nu} z_\nu.
\]
As previously discussed, the closure under time differentiation of
the space spanned by the first order kernels is assured by
${\bf\dot{z}}$, as in Eq.~\eqref{eq:depsilon} in the appendix.  Note
that our separation of $W[{\bf j}]$ depends on this particular
property. For a fixed spike time $t$, restricting ourselves to the
region where $\tau < t$, $W_\perp [ \jj ]$ becomes a static
integral. Thus Eq.~\eqref{eq:static1} shows clearly how the
properties of this model emerge. $P\cnd{\spike}{{\bf s}}$ is just
given by a threshold with a weight function $w(\zz)$ up to some
linear transformations. Therefore, for example, the linear threshold
case reduces to the one dimensional case. Since $\nabla_\zz \theta (
\zz )$ is a constant vector, every computation reduces to a one
dimensional integral in this direction and the corresponding filter
is a linear combination of $f_\mu$s, as we have observed in
Fig~\ref{fig:FNfirst}d. 
Furthermore, a
variety of structures may be generated depending on how trajectories
cross the threshold.

From Eqs \eqref{eq:static1} and \eqref{eq:static2}, we can derive
the analytic forms of the first and second order moments:
\begin{equation}\label{eq:staticSTA}
\text{STA} ( \tau ) = \sum_{\mu} c_{\mu} f_{\mu} ( \tau ), \qquad c_{\mu}
        = \frac1N \int d^n \zz \, z_{\mu} e^{- z^2 / 2\sigma^2} w ( \zz ),
     \quad N = \int d^n \zz\,e^{-z^2/2\sigma^2}w(\zz),
\end{equation}
and
\begin{equation}\label{eq:staticCov}
 C ( \tau, \tau' ) = \sum_{\mu,\nu} \left(c_{\mu \nu}- c_\mu c_\nu
 -\sigma^2 \delta_{\mu\nu} \right) f_{\mu} ( \tau ) f_{\nu} ( \tau' ),
 \quad c_{\mu \nu} = \frac1N \int d^n\zz\,z_{\mu} z_{\nu} e^{- z^2 / 2\sigma^2} w ( \zz ).
\end{equation}

Note that we have assumed a prior based on the distribution of
randomly driven trajectories of the linear system. This takes no
account of perturbations of this distribution due to flux from the
system's return from a spike. This assumption is equivalent to
assuming that the last spike is in the distant past, so that memory
of that perturbation has vanished. In this paper we will consider
only this ``isolated spike" case. We will return to this point in
the discussion.

\subsection{Variance dependence}\label{sec:variance-dependence}

Through Eqs. \eqref{eq:staticSTA} and \eqref{eq:staticCov}, this
model captures an explicit dependence of the white noise outcome on
the stimulus statistics, in particular, the variance $\sigma^2$.
While the subthreshold dynamics are linear, the dependence on
$\sigma$ is nonlinear due to the threshold shape and the weight
function $w(z)$. We discuss two examples below.

\begin{figure}[tp]
        \begin{center}
            \includegraphics[scale=1]{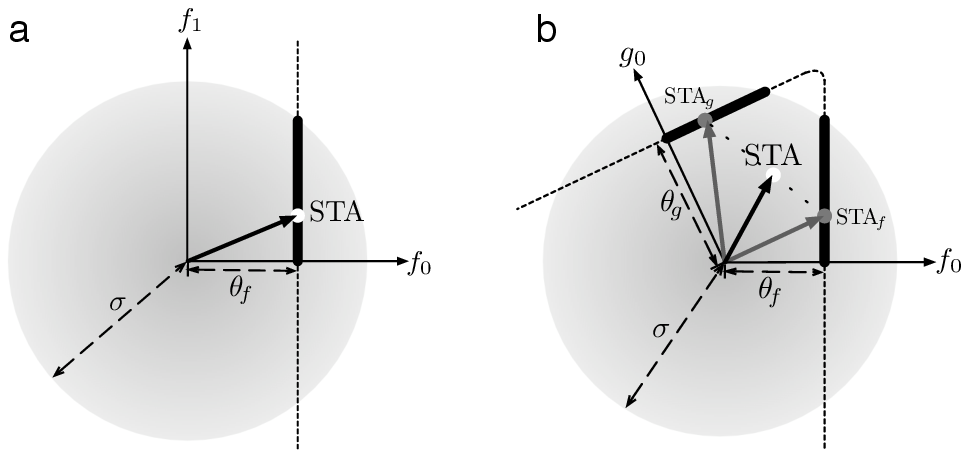}
    \end{center}
    \caption{(a) A diagram of a single filter model. The disc of radius $\sigma$
    represents the prior Gaussian distribution. The dotted
    line is the model threshold imposed at $z_0 = \theta_f$. The
    thick line is a distribution of spike triggered stimuli, and the white dot
    represents its average, the $\STA$. (b) A diagram of a model whose threshold
    is the union of two lines, each imposed at a distance $\theta_{f,g}$ from the
    center. Two gray dots denote the STAs for each segment.}
    \label{fig:filter-diagram}
\end{figure}

We first consider a linear threshold. Through a suitable linear
transformation, this case simply reduces to a filter-and-fire model
with a single filter, say $f_0$, and a fixed threshold $z_0 =
\theta_f$. Now the distribution of threshold crossing points is
constrained by $z_0 = \theta_f$ and $\dot{z}_0
>0$. As we mentioned in the previous section, $\dot{z}_0$ lies in
the originally defined feature space, and is given by another single
filter, which we denote by $f_1$. In other words, when the
normalized $\dot{f}_0$ is denoted by $f'_0$, we can choose $f_1 =
f'_0$ by a suitable orthogonal linear transformation. Thus, our
system depends on two filters, which are depicted schematically in
Fig~\ref{fig:filter-diagram}a.

The STA is the centroid of the distribution of threshold crossing
points. Eq.~\eqref{eq:staticSTA} reproduces a previously known
result~[Meister, M., private communication]\cite{fairhallRetina},
\begin{equation}\label{eq:STA-single-threshold}
    \STA(\tau) = \theta_f f_0(\tau) +
    \frac{\sigma}{\sqrt{\pi/2}} f_1(\tau).
\end{equation}
With high variance $\sigma\gg\theta_f$, the STA is dominated by
$f_1$.  In the feature space, with increasing variance the threshold
stays the same, but a larger portion of it is crossed by
trajectories driven by the larger variance ensemble, as can be seen
in Fig~\ref{fig:filter-diagram}a.

When the threshold is curved, the $\sigma$ dependence is considerably more
complicated. We will consider an extreme but analytically tractable version of
this case to illustrate the point: let the curved threshold be approximated by
two linear ones imposed at $\theta_f$ and $\theta_g$ in the $ f_0$ and $g_0$
directions respectively, as in Fig~\ref{fig:filter-diagram}b. In this
case, one segment of the threshold imposes a dependence on the filters $f_0$
and its (normalized) derivative, $f_1$, while the other selects $g_0$ and
$g_1=\dot{g}_0/\|{\dot{g}_0}\|$. The space of relevant features is still two
dimensional, or three including the $\delta$-function, since $g_0$ and $g_1$
are linear combinations of $f_0$, $f_1$ and possibly also $\delta(t)$.

Now the STA of this system is
\begin{equation}\label{eq:STA-double-threshold}
    \STA(\tau) = \cos^2\varphi\cdot\STA_f(\tau) +
     \sin^2\varphi\cdot\STA_g(\tau),\quad \varphi =
    \tan^{-1}e^{(\theta_f^2-\theta_g^2)/4\sigma^2},
\end{equation}
where $\STA_\epsilon(\tau)= \theta_\epsilon \epsilon_0(\tau) +
\frac{\sigma}{\sqrt{\pi/2}} \epsilon_1(\tau)$ as in
Eq.~\eqref{eq:STA-single-threshold}. Note that as in
Fig~\ref{fig:filter-diagram}b, the STA does not lie on the
threshold. As for a variety of experimental examples such as complex
cells \cite{touryan}, neurons of rat barrel cortex
\cite{rasAbstract}, and some retinal ganglion cells
\cite{fairhallRetina}, the spike-triggered stimuli are poorly
represented by their first order statistics, the STA. Also, the
coefficients of $\STA_{f,g}$ depend exponentially on $\theta_{f,g}$
and $\sigma$. Thus, the system shows a nonlinear dependence on the
stimulus variance.

\begin{figure}[p]\centering
    \includegraphics[scale=1]{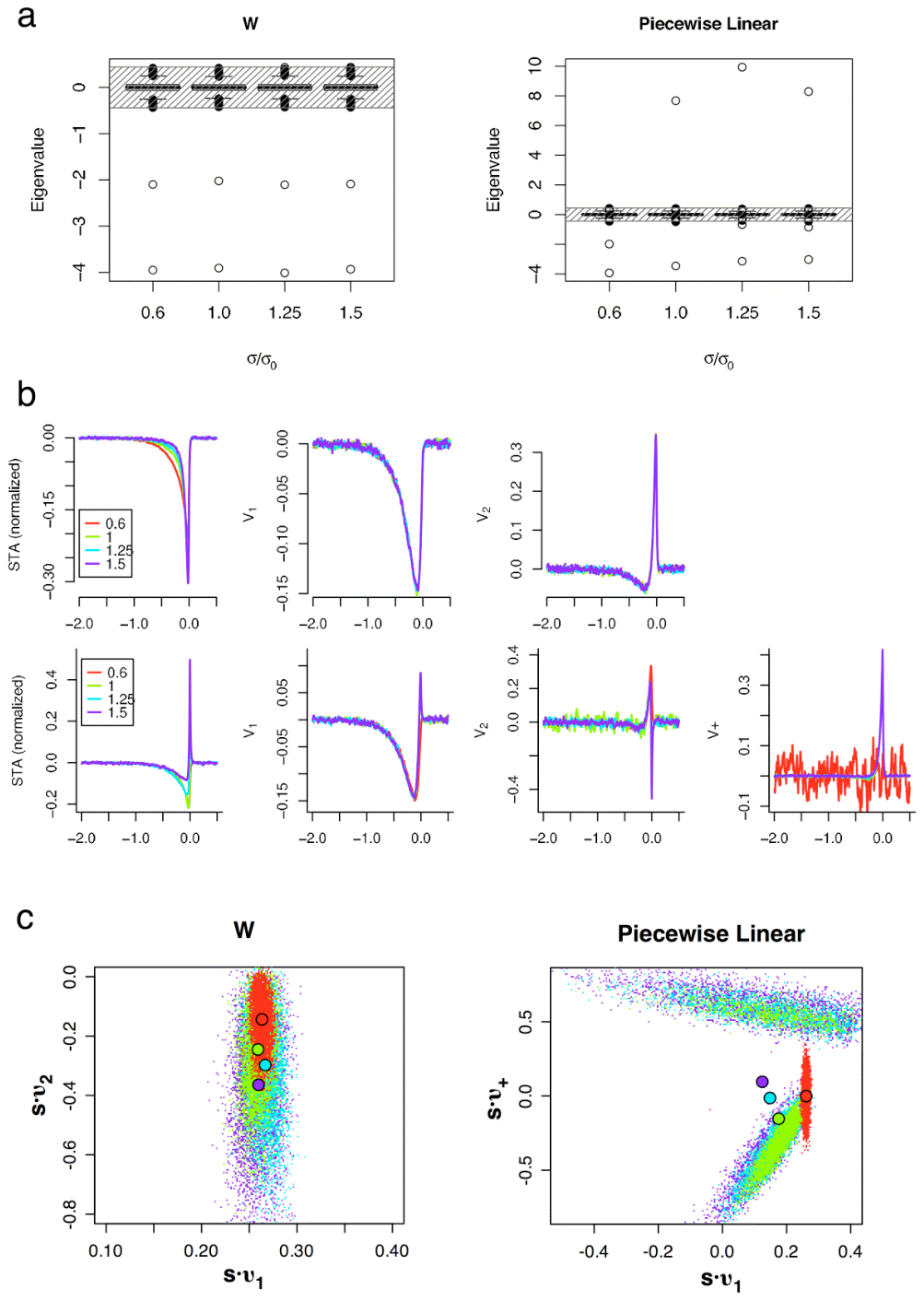}
    \caption{Two examples of variance dependence in the linear system
    model.
(a) Spectra of the covariance matrix for several different variance
stimuli. (b) Significant eigenmodes. (c) Projection of
spike-triggering stimuli onto the filter subspace defined by
$v_{1,2}$ for the linear and $v_{1,+}$ for the piecewise linear
threshold respectively, colored according to the stimulus variance
as in the legend of (b). The circled points are the centers of the
distributions, corresponding to the respective STAs, colored
according to the respective distribution.}
    \label{fig:linear-vs-pl}
\end{figure}

Fig~\ref{fig:linear-vs-pl} shows an example. The $W$ threshold model
does not show any significant change, except sharpening of the STA
due to the increased component of $\dot{\epsilon}_W$,
Fig~\ref{fig:linear-vs-pl}b, and linear broadening of the spike
triggered stimulus distribution. The piecewise linear threshold case
is more dramatic: while the smallest variance does not drive the
system hard enough to produce a positive eigenmode (hence the
one-dimensional distribution of projections seen in
Fig.~\ref{fig:linear-vs-pl}b and c, red), a new significant mode
emerges as the variance increases. The STA changes beyond sharpening
and almost looks like a different model at high variance compared
with low variance. Each of the significant modes change as more
trajectories cross the threshold from the other side and the
principal axis of the distribution rotates, as seen in
Fig~\ref{fig:linear-vs-pl}c.

\section{Dynamical threshold}\label{sec:dynamical-threshold}

In the previous section, we considered neuron models composed of
linear filters derived from the FN neuron, and some
choices of imposed thresholds. However, when we consider the full FN
neuron, the threshold arises from the structure of the FN equations
\eqref{eq:fhna}-\eqref{eq:fhnb}. Here we discuss the importance of
the threshold identification for reverse correlation analysis.

\begin{figure}[p]
    \begin{center}
        \includegraphics[scale=1]{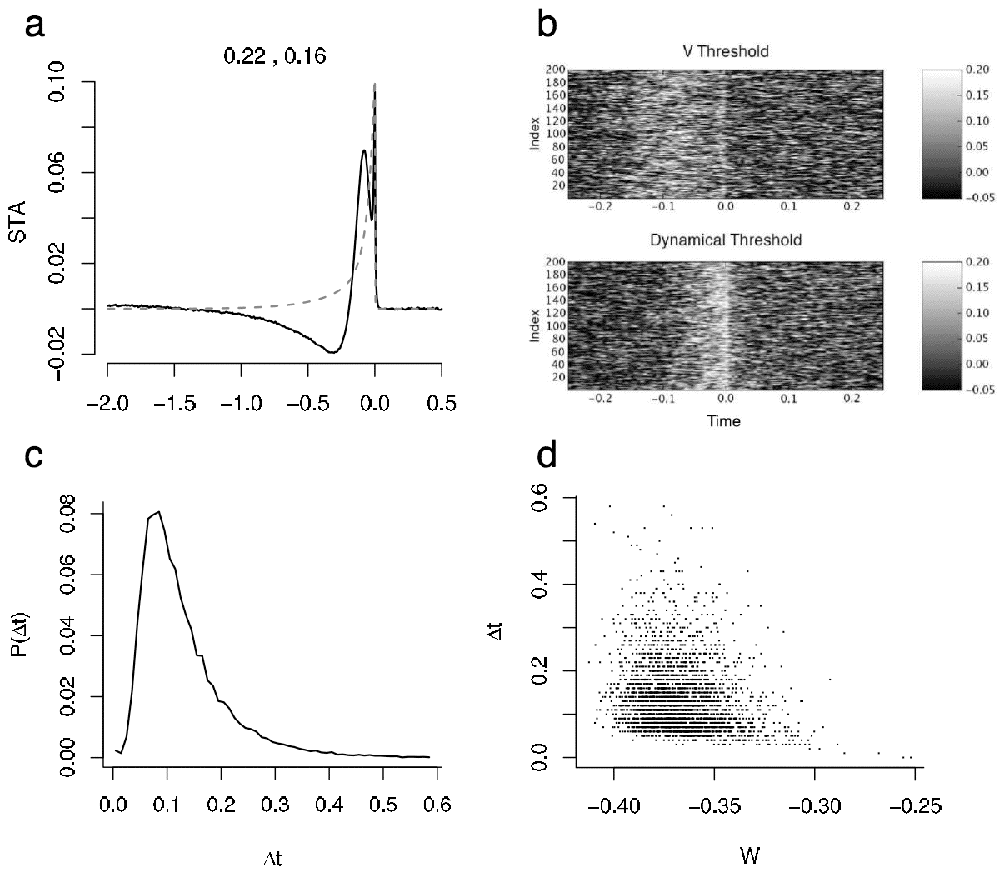}    
\end{center}
\caption{Effect of a dynamical
threshold in the FN model.
(a) STA of the full FN system with a constant $V$
  threshold. As before, a linear fit by the kernels is also shown.
(b) Comparison of spike triggered stimuli with different
  threshold choices in the FN model. (c) Temporal point spread function
induced by the choice of threshold. $\Delta t$ is the time delay from
crossing the dynamical threshold to the fixed $V$ threshold.
(d) Time delay of each spike
triggered stimulus plotted against the value of $W$ at its threshold
crossing point. }
\label{fig:FNalign}
\end{figure}

The problem of the identification of a threshold arises immediately
upon attempting a reverse correlation analysis; spike times are
often defined by the threshold crossing of the voltage. Here also,
we can impose an arbitrary threshold in $V$ to identify each spike.
The STA obtained using this scheme is displayed in
Fig~\ref{fig:FNalign}a. It is clear that it cannot be well described
by the first order kernels. However, this does not mean breakdown of
the analysis; rather, it underscores the point that the STA or any
other single spike quantities should be computed using the ``correct''
threshold that we take to be the dynamical threshold discussed in
\S\ref{sect:intro}.

Fig~\ref{fig:FNalign}b compares the spike triggered stimuli in the
fixed $V$ and dynamical threshold cases. While the peaks of the
stimuli are spread out in time in the fixed $V$ threshold case, for
the dynamical threshold the peaks lie in a narrow band around the
spike time. As we can see from Fig~\ref{fig:FNphaseplane}, the
spiking trajectories cross the dynamical threshold before crossing
the $V$ threshold, and this timing difference, $\Delta t$, depends
on $W$. Additionally, since the system is driven by Gaussian white
noise, the variability increases with the timing difference,
inducing a point spread function on the STA. Hence, each spike
triggered stimulus is contaminated by this temporal jitter, and
estimated filters are distorted. As mentioned in
\S\ref{sec:minimal-spiking-models}, the FN neuron does not have a
clear-cut dynamical threshold; the threshold was chosen to some
degree arbitrarily, and this will induce some error in the
estimation of filters.  However, Fig~\ref{fig:FNstav}a shows the
improvement in the STA computed using the dynamical threshold.
Figs~\ref{fig:FNalign}c and d show the point spread
distribution -- the distribution of values of $\Delta t$ -- and how it
is correlated with $W$. This situation is similar to that discussed
in \cite{zane}, where it was noted that temporal or spatial jitter
can blur the estimation of filters and receptor fields. There, a
blind deconvolution algorithm was used to ``dejitter'' and realign
the spike-triggered stimuli, which dramatically sharpened the
estimate of the spike-triggering stimulus. In a real system, jitter
may indeed be due at least partly to noise, but it is also possible
that a component of such jitter is deterministic and due to
variability in the point of dynamical threshold crossing, as in the
FN case. Blind deconvolution may then be viewed as an empirical
approach to recovering a dynamical threshold based on spike times
originally recorded using a voltage threshold. It is interesting to
note that estimated jitter in such a case is correlated with
projection onto the STA derivative.\footnote{This correlation arises
because the STA derivative is the linear approximation to time
translation of the STA.  Hence, if a current history aligns with the
STA better under a small time translation or jitter, it will have a
projection onto the STA derivative proportional to that jitter.  In
Sect. 2 of this paper it was demonstrated that the STA and its
derivatives are, up to a linear transformation, the filters
associated with the dynamical variables of the linearized system.
Therefore there is a relationship between blind deconvolution to
optimize fit to an STA and estimation of a dynamical threshold based
on the ``hidden'' (non-$V$) dynamical variables.}

\section{Fully nonlinear systems}

In this section, we discuss the covariance analysis of the full FN
system and compare the result with the same analysis of the first
and second order approximation.

We identified $\sim 2\times10^6$ spikes first using a voltage
threshold, and then backtraced each trajectory to the point where it
crosses the dynamical threshold shown in Fig~\ref{fig:FNphaseplane}.
The trajectory might cross the threshold multiple times before
spiking due to the noisy input. In this case, we used the first
crossing point after the trajectory diverges from that of the second
order approximation. This is based on the assumption that the second
order is a sufficiently good approximation of the system in the
subthreshold regime. For both the first and second order system,
crossing of the dynamical threshold is used to identify spikes.
Right after a spike, we imposed a post-spike inhibitory period equal
to about a ``spike width'', as empirically determined from the full
system.

\begin{figure}[p]\centering
    \includegraphics[scale=1]{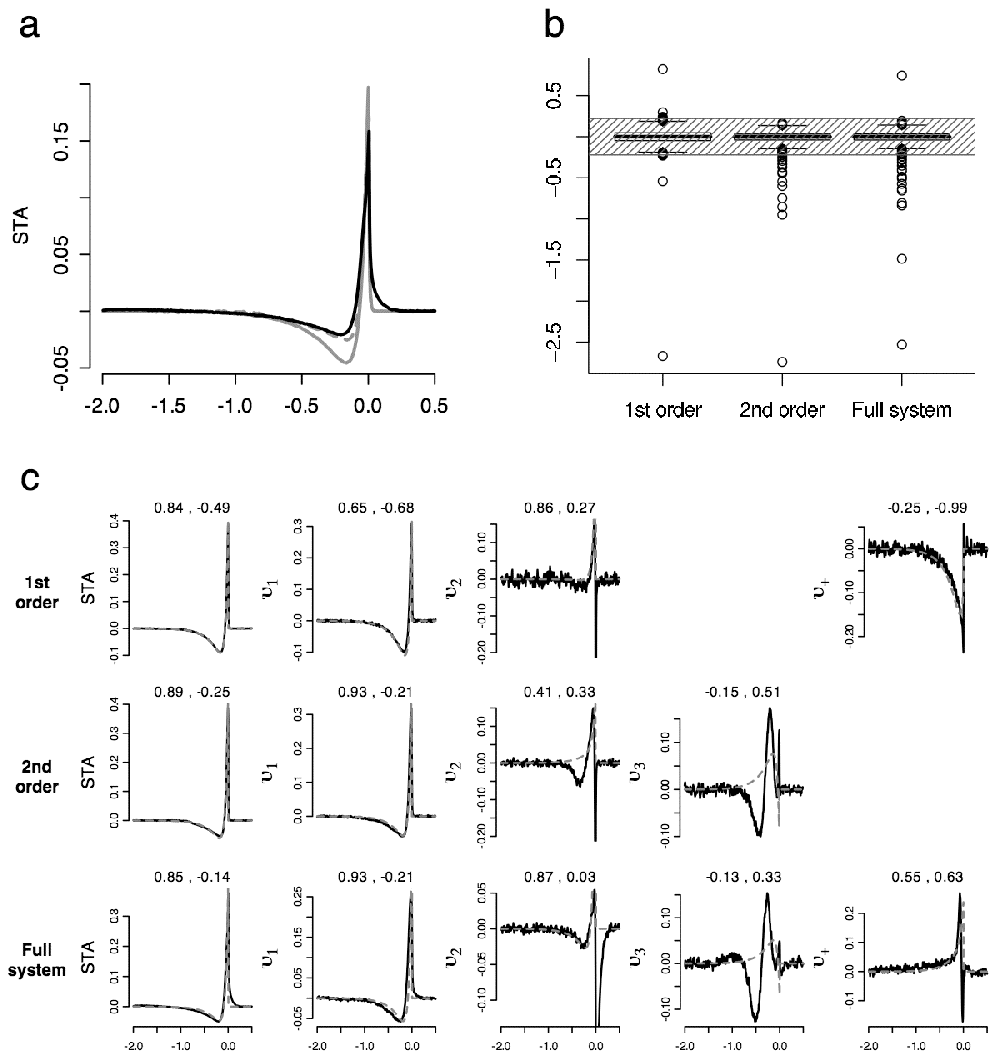}
\caption{ (a) STAs of a FN model neuron, and the first (gray) and second (dashed) order approximation.(b) Spectra of covariance matrices of the FN model and
approximations. (c) STAs and covariance
modes for the first and second order approximations and the full FN
system.}
\label{fig:FNstav}
\end{figure}

\begin{figure}[p]\centering
\includegraphics[scale=1]{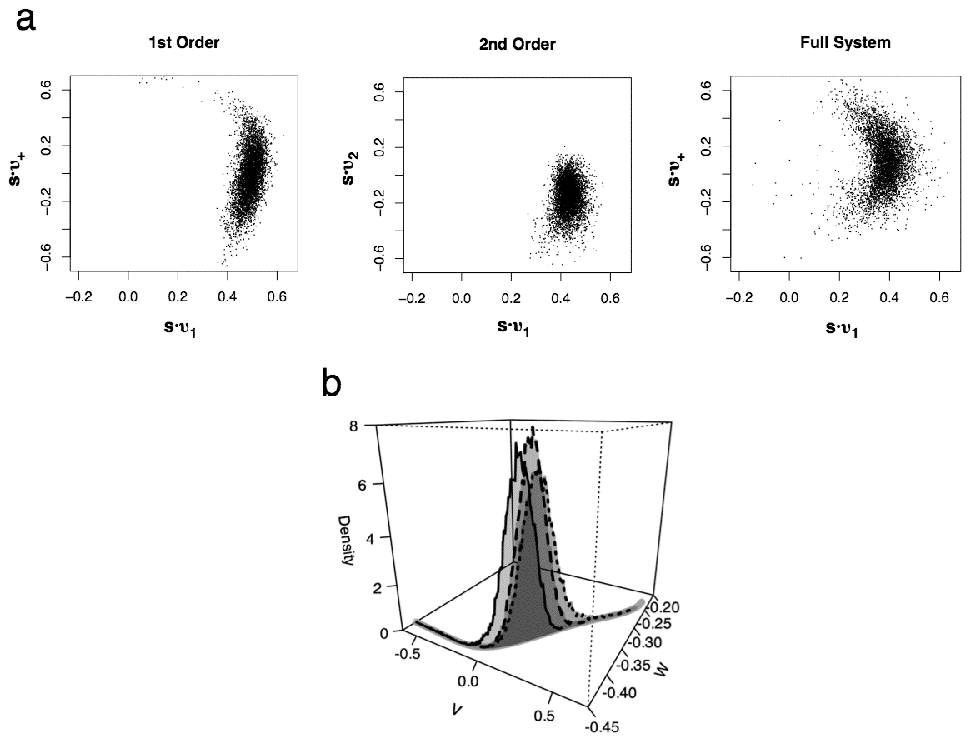}
\caption{(a) Spike triggered stimuli projected into a feature space
defined by $v_{1,+}$ for the first order case and full system, and
$v_{1,2}$ for the second order system. (b) The density of threshold
crossing points in the first order (solid), second order (dashed), and
full (dotted) systems, plotted along the threshold curves in the $V-W$
plane. }
\label{fig:FNproj}
\end{figure}

The results are shown in Fig~\ref{fig:FNstav} and~\ref{fig:FNproj}.
Fig~\ref{fig:FNstav}a shows that the STA of the first order
approximation is quite similar to that of the full system, and the
second order STA is even closer. From the covariance analysis, we
see that the FN neuron, like the HH model~\cite{hh}, has one
dominant negative and one dominant positive eigenvalue.  However,
the results from the covariance analysis of the three systems differ
considerably. The second order and the full system both show a
relatively large number of significant eigenvalues. This is due to
the contribution of the second $g_2[V]$ and $g_2[W]$ (and higher
order) kernels in Eq~\eqref{eq:2ndwieneriny}. However, these are
relatively suppressed compared to the first order modes. Further,
the spectrum derived from the second order system,
Fig~\ref{fig:FNstav}b, has no positive significant eigenvalue.

Fig~\ref{fig:FNstav}c provides more detailed information. The first
order case is just as we have seen previously, with three modes
which are well described by linear combinations of the linear
kernels. In the second order case, $v_1$ and $v_2$ are comparable to
those of the first order. $v_3$, which is not approximated by the
first order kernels, must arise from the second order kernels. In
this regard, the full system is well matched with the second order
approximation. There also, $v_3$ is comparable to that in the second
order, and $v_{1,2}$ are derived from the linear kernels. However,
the full system also has a positive mode, $v_{+}$, from the linear
kernels, which resembles the first order rather than the second.

The geometry of the spike-triggering stimulus projections is shown
in Fig~\ref{fig:FNproj}a. The first order, not surprisingly,
recapitulates the curved threshold case in
Fig~\ref{fig:FNfirst}d. However, the second order is more like
the $(V,W)$-linear threshold case, while again the full system
resembles the curved in Fig~\ref{fig:FNfirst}d.  A possible
explanation for this is that each system probes different parts of
the dynamical threshold. Fig~\ref{fig:FNproj}b marks the
density of threshold crossing points for each model. In contrast to
the first order and the full system, which access a large section of
the threshold with non-trivial curvature, the second order only
probes a small and almost linear section. This is the reason for the
lack of a positive eigenvalue. As in the toy models in
\S\ref{sect:linear}, the contributions of the relevant dimensions
are determined not only by local information (filters) but also by
the global structure of a multidimensional or dynamical threshold.

\section{Abbott-Kepler model}\label{sec:abbott-kepler-model}
In this section, we apply the same analysis to a two-dimensional
model which is more nonlinear and more realistic than the FN model.

Abbott and Kepler \cite{abbott&kepler} developed a two dimensional
reduction of the Hodgkin-Huxley model, based upon the observation
that there is a separation of timescales between the faster $m$ and
the slower $n$ and $h$ variables.  $m$ is then replaced with its
asymptotic value at the membrane voltage $V$, while $n$ and $h$ are
controlled by another voltage variable $U$. The equations of the
Abbott-Kepler (AK) model are of the form
\begin{eqnarray}
        C\frac{dV}{dt} & = & f(V,U) + I(t),\label{eq:ak1}\\
        \frac{dU}{dt} & = & g(V,U),\label{eq:ak2}
\end{eqnarray}
where $f(V,U)$ and $g(V,U)$ are nonlinear functions in $V$ and $U$.
Their derivation is briefly sketched in appendix~\ref{sec:derivation-abbott-kepler}, and we refer to the original paper \cite{abbott&kepler} for further detail.
The nullcline for $U$ is given by $g(V,U) = 0$, which is satisfied
by $V=U$. The $V$ nullcline, $f(V,U)=0$, is more complicated and
obtained numerically. The two nullclines intersect at a fixed point
$V=U=-65\text{mV}$.

\begin{figure}[tp]
    \begin{center}
        \includegraphics[scale=1]{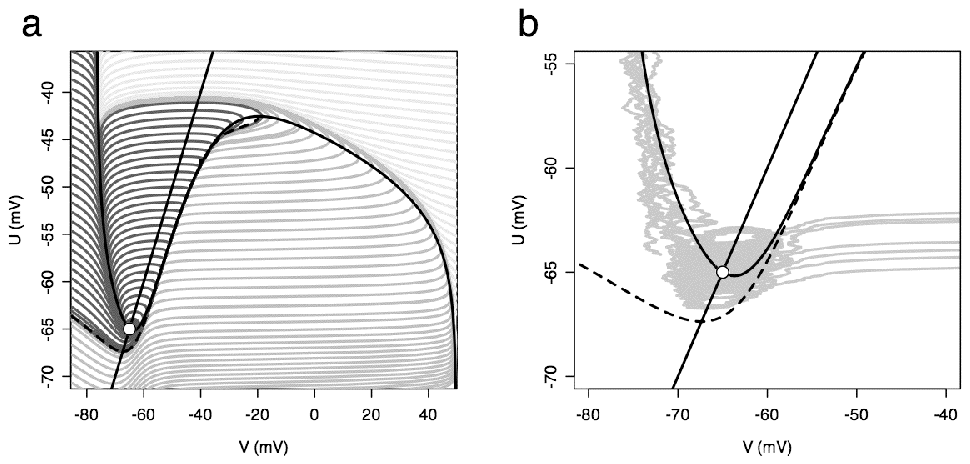}
    \end{center}
    \caption{(a) Phase plane of the Abbott-Kepler model with trajectories,
    nullclines, and a threshold, in the absence of input. (b) AK phase plane
    with the injected noisy input.}
    \label{fig:AKphaseplane}
\end{figure}

\begin{figure}[p]
    \begin{center}
        \includegraphics[scale=1]{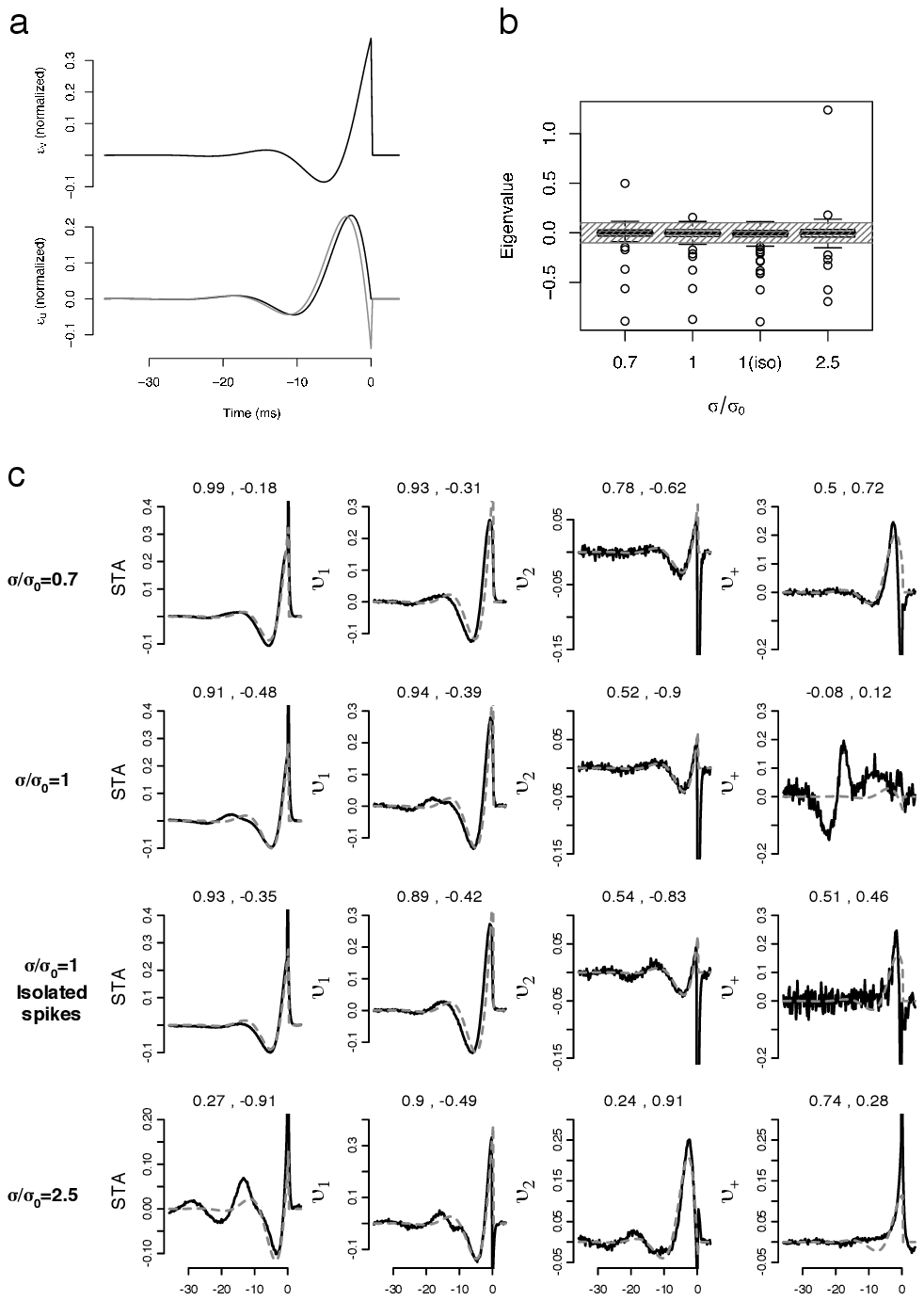}
        \end{center}
\caption{(a) Normalized first order kernels $\bar{\epsilon}_{V,U}$ of the AK
model and the component of $\epsilon_U$ orthogonal to $\epsilon_V$, normalized
(gray). (b) Eigenvalue spectrum of the covariance matrix of the AK model with
each variance. $\sigma_0 = 13\text{pA}.$ (c) Eigenmodes of the covariance
matrices. As previously, $v_{1,2}$ and $v_{+}$ correspond to the two smallest
negative eigenvalues and the largest positive one.}
    \label{fig:akresult}
\end{figure}

Fig~\ref{fig:AKphaseplane}a shows the phase plane of this model with
zero input current. Like Fig~\ref{fig:FNphaseplane}, the threshold
structure, which can be obtained numerically, is visible. Due to the
strong nonlinearity, spiking trajectories are well-defined on the
phase plane and the threshold has less ambiguity than for the FN
model. Again, we try to identify the dynamics of the system in the
subthreshold regime with the first order approximation. Unlike a FN
neuron, the Jacobian of this system has complex eigenvalues
$\lambda_\pm = -0.2118 \pm i0.4035 \text{ms}^{-1}$, and therefore
the first order kernels $\epsilon_{V,U}^{(1)}$ oscillate,
Fig~\ref{fig:akresult}a. This is consistent with the oscillatory
linearized behavior associated with the full Hodgkin-Huxley model
near equilibrium.

\begin{figure}[tp]
    \begin{center}
        \includegraphics[scale=1]{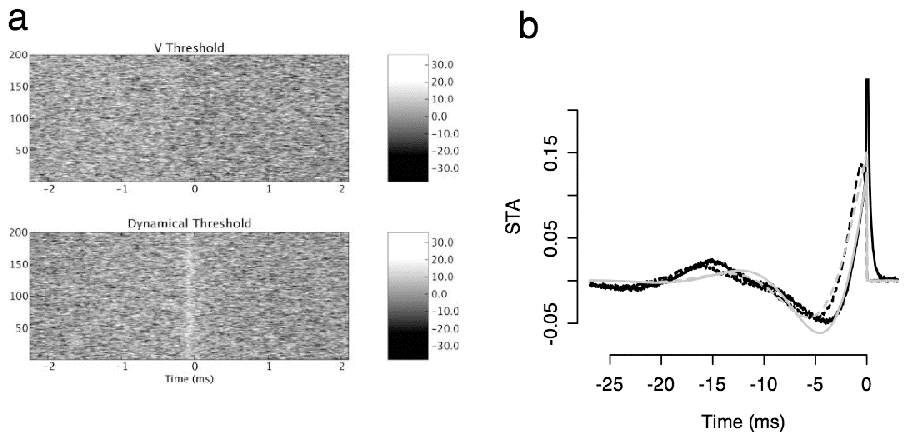}
    \end{center}
    \caption{(a) Spike triggering current histories with two different
    threshold schemes in the AK model. (b) STAs of the AK model with the fixed $V=-40\text{mV}$ threshold
    (dashed) and dynamical threshold (solid). Also, they are
    least-squares
    fitted by linear combinations of $\epsilon_{V,U}$ (gray).}
    \label{fig:AKalign}
\end{figure}

Before we carry out covariance analysis on this model, we examine
the effect of a dynamical threshold, as in
section~\ref{sec:dynamical-threshold}. Fig~\ref{fig:AKalign}a shows
the spike triggering stimuli aligned both with a fixed threshold in
$V$, chosen as $V=-40\text{mV}$ to unambiguously select spiking
trajectories, and the dynamical threshold. The typical time shift is
of order $\leq1\text{ms}$; the overall STA only suffers from a
slight time/phase shift when the $V$ threshold is used. However, in
the small time scale of $\leq1\text{ms}$, there is discrepancy from
the dynamical threshold case, which is characterized by a large
delta-function component at the spike time. Both STAs are well
approximated by linear combinations of $\epsilon_{V,U}^{(1)}$, up to
a small deformation around -15ms due to the effects of multiple
spikes.

\begin{figure}[p]
        \begin{center}
            \includegraphics[scale=1]{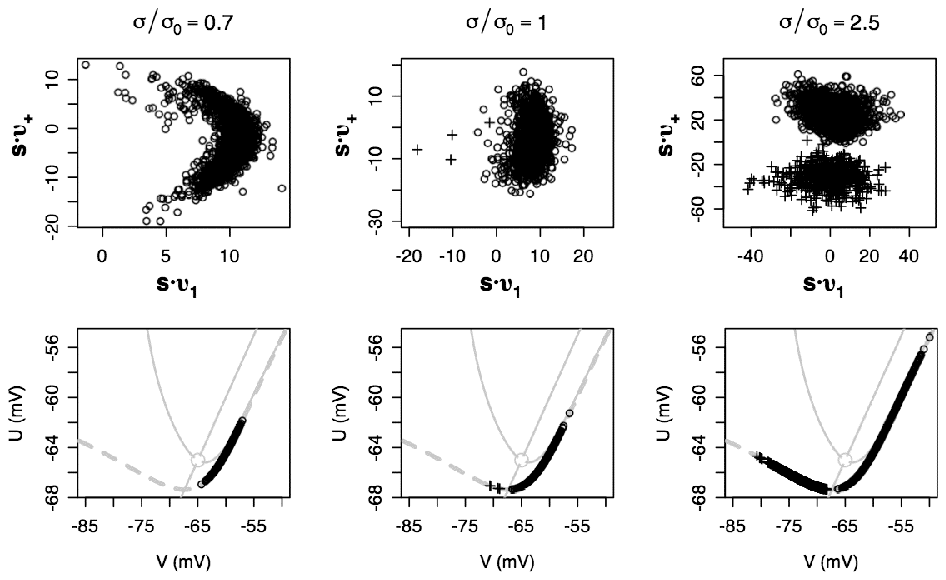}
        \end{center}
    \caption{
    Projection of spike triggering stimuli onto $v_1$ and $v_{+}$
    in the AK model.
    Below are threshold crossing points in the phase plane of each case,
    marked by crosses, $V\leq-67\text{mV}$, and circles, when
    $V\geq-67\text{mV}$, over the dynamical threshold (dashed).
    The other gray curves are the nullclines,
    with their intersection, the fixed point, marked by a circle.
    }
    \label{fig:akprojcrossing}
\end{figure}

Fig~\ref{fig:akresult} shows the results from covariance analysis
carried out with $\sim2\times10^6$ spikes for Gaussian white noise
stimuli with various variances. For comparison with previous
results, we selected three eigenmodes corresponding to the leading
two negative and the largest positive eigenvalues. We find that they
are reasonably well approximated by linear combinations of
$\epsilon_{V,U}^{(1)}$ in most cases, although they are sometimes
affected by a $\delta$-function at $t=0$ and a large multi-spike
effect at high variances.  The multi-spike effect can be eliminated
by considering only the isolated spikes when the spike rate is
low~\cite{i&f}. At higher variances isolated spikes are rare and
there is a stronger influence of oscillating ``silence
modes'' \cite{hh}. As in the FN neuron case, we identify modes
other than those in Fig~\ref{fig:akresult}c as ``nonlinear modes''.

We compare results at different variances with our discussion in
\S\ref{sec:variance-dependence}. Some features of variance
dependence are shared with the toy model in
\S\ref{sec:variance-dependence}: the eigenvalue spectrum drifts and
the corresponding modes rotate among themselves. However, the
Abbott-Kepler modes also exhibit more complicated behavior.
Fig~\ref{fig:akprojcrossing} shows projections of spike triggered
stimuli onto two distinguished eigenvectors $v_{1,+}$ and
corresponding threshold crossing points. At low variance, the system
crosses mostly one side of the threshold while the projections trace
out the curvature of the threshold segment. As the variance
increases, some crossing points begin to appear on the left side of
the threshold. However, this does not overcome the expansion of a
crossing point distribution on the right side, and the modes
corresponding to this direction dominate. At high variance, there
are many crossing points on both sides, reflected in the bimodal
distribution of the projections, previously seen in the toy models.
Thus this is another example of how the results of reverse
correlation analysis are affected both by the filter properties and
by the interaction of the stimulus ensemble with the threshold
geometry.

So far, we have discussed only two dimensional dynamical models. We
began with some artificial toy models with purely linear
subthreshold dynamics, and proceeded to the minimal FitzHugh-Nagumo
spiking model. We applied the lessons learned there to the more
nonlinear Abbott-Kepler model. We will conclude with a partial
analysis of the higher dimensional Hodgkin-Huxley model.

\section{Hodgkin-Huxley model}

Higher dimensional systems require nontrivial extensions of the
methodology we have used with two dimensional systems. First, it is
much harder to use the phase portrait to find a dynamical threshold,
which could now be a multi-dimensional hypersurface rather than a
curve. If we do not align the spike triggered stimuli according to
the dynamical threshold, the obtained filters may include a
distribution of time delays. For example, the reverse correlation
analysis may result in broadened filters,
\begin{equation}
\tilde{f}_\mu(t) = \int d\tau\, f_\mu(\tau) p_\sigma(t-\tau),
\end{equation}
where $p_\sigma(\tau)$ is a point spread function, depending on
the input variance $\sigma^2$, as in
Fig~\ref{fig:FNalign}c.\footnote{In fact, it can be much more
complicated than this. Since the time delay depends on the location
of threshold crossing in a phase space, additional temporal
correlations are introduced. Therefore, each filter may have a
different point spread function, or filters may depend upon the time
delay structure in a complicated way. We will discuss here only the
simplest case.}

However, we can still gain some insights from $\tilde{f}_\mu$. If
$p(t)$ has a narrow support, say less than a millisecond, it has
negligible effects for larger time windows. It is also possible that
some consequences of our analysis will still apply to
$\tilde{f}_\mu$ with only a few reasonable assumptions. For example,
if $p_\sigma$ is of limited temporal extent, then the derivatives of
$\tilde{f}_\mu$ can be simply
\begin{equation}
    \tilde{f}^{(n)}_\mu \approx \int d\tau\, f^{(n)}_\mu(\tau) p_\sigma(t-\tau).
\end{equation}
Therefore, as we discussed in \S\ref{sect:linear}, the linear modes
among $\{\tilde{f}_\mu\}$ should still be (approximately) closed
under time differentiation.

\begin{figure}[p]
        \begin{center}
\includegraphics[scale=1]{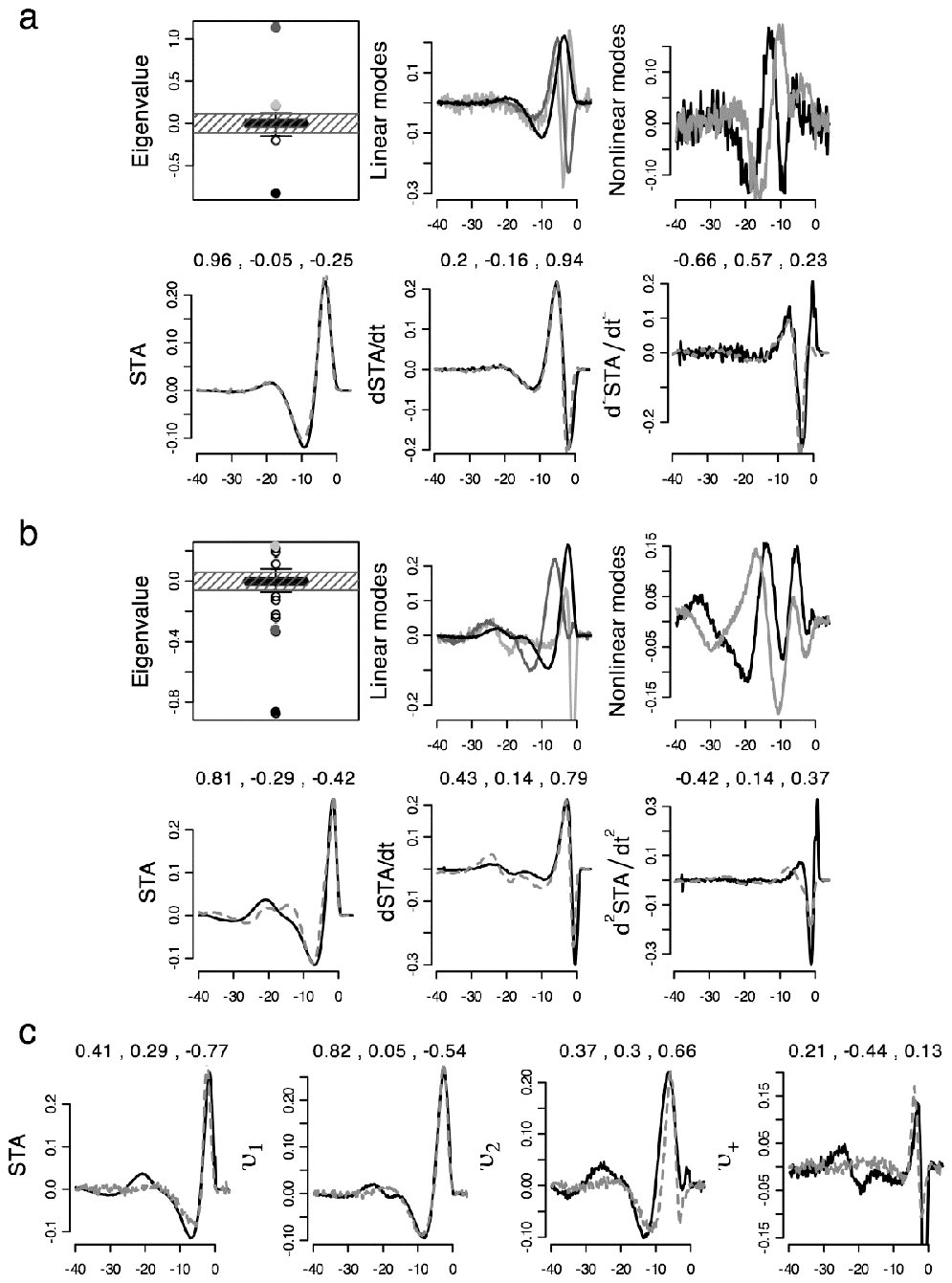}
    \end{center}
    \caption{Covariance bases of the linear feature space of the Hodgkin-Huxley
    model recovered at (a) low and (b) high variance. The three most significant eigenvalues are
    denoted with the same colors as the corresponding linear eigenmodes. (c) The STA and
    linear modes of the high variance case and their fits using the linear
    modes of the low variance case in the HH model.}
    \label{fig:HH}
\end{figure}

To demonstrate this, we use the four dimensional Hodgkin-Huxley model,
and the following strategy: we select the linear modes out of the
significant filters from covariance analysis. Now, since the STA is
approximately their linear combination, the time derivatives of the
STA should be written in terms of the linear combinations as well if
they are closed under time differentiation. Fig~\ref{fig:HH} shows
that this holds. In the low variance case, Fig~\ref{fig:HH}a, the
three linear modes chosen provide good fits to the time derivatives
of the STA. The high variance case, Fig~\ref{fig:HH}b, is
affected by multi-spike effects and a filtering artifact, but it is
also well fitted. Therefore, we can conclude that the time
derivatives of the STA span the same space as three linear modes.
One might expect a fourth linear mode due to the dimensionality of
the model, but this is not as significant as the others; this agrees
with previous covariance analysis that the HH model can be well
described as a quasi-three dimensional system \cite{hh}.

We note that it is not clear whether the linear modes in the two
cases span the same feature space. This is difficult to ascertain
since the point spread function can in principle depend on the
stimulus variance. In Fig~\ref{fig:HH}c, we see that while
$v_1$ of the high variance case can be fitted by the linear modes of
the low variance, the other modes show a small deviation even around
5ms. This might indicate an interesting variance dependence as in
\S\ref{sec:variance-dependence}, but we will not pursue this issue
in this paper.

\section{Summary}

Here we have investigated the meaning or interpretation of the
features derived from the spike-triggered covariance method applied
to two simple neuron models, the FitzHugh-Nagumo model, the minimal
spiking neuron model, and the Abbott-Kepler model, a more faithful
two-dimensional reduction of the Hodgkin-Huxley system. The power of
white noise analysis is that it provides a data-driven method to
reduce a high dimensional dynamical system to a functional model
which captures the essential computation of the system in terms
familiar to systems neuroscience: a receptive field which filters
the stimulus, and a threshold function over the filtered stimulus.
Our goal here was to analyze the output of a white noise analysis in
terms of what it can reveal and how it depends upon the underlying
dynamical system. In this, our approach is distinct from the elegant
work of \cite{huysPaninski} where responses to white noise stimuli
are used to fit the parameters of a conductance-based model.

Dealing with simple two-dimensional systems, our observations of
spiking dynamics in the phase plane motivated the following reduced
model:  dynamics in the phase plane are approximated by the
perturbative expansion, in particular the linear approximation, and
the system's nonlinearity is captured by a spiking threshold,
determined for zero input, that extends through multiple dimensions.
This model is a generalization of a basic filter-and-fire model to
multiple dimensions, and extends it in two significant ways. One is
the identification of the filters with the dynamical variables of
the system; the other is the generalization of the concept of the
threshold.

The simplified model with linear dynamics and a multidimensional
curved threshold was treated both analytically and
phenomenologically, using numerical simulation and reverse
correlation (covariance) analysis of spike triggered stimuli. This
led to several insights, some of which are related to existing
observations. First, for this case the feature space derived from
covariance analysis is spanned by the linear kernels. Since the set
of kernels is closed under time differentiation, the same feature
space can be also spanned by a generic linear combination of the
kernels, such as the spike triggered average and its $n$ time
derivatives. However, not every kernel contributes as a relevant
feature.  The threshold structure plays a role of selecting the
relevant ones from the kernels: for example, a linear threshold
selects a single filter and its time derivative. In general, a
threshold spanning a $d$ dimensional subspace will select $d+1$
features from among linear combinations of the kernels. We show
further for this model that the distribution of spike triggered
stimuli in the covariance feature space corresponds to that of
threshold crossing points, up to a suitable linear transformation.

We also used this model to illustrate the effects of the interaction
of the threshold nonlinearity with the stimulus ensemble. We show
that a complex threshold geometry leads to a nontrivial variance
dependence of the eigenmodes and eigenvalues of covariance analysis,
and even more so, the spike-triggered average. In the linearized
subthreshold model, the subspace of eigenmodes is not changed, but
the spike-triggering ensemble may rotate through this subspace,
leading to a variance-dependent spike-triggered average. This is one
example of stimulus variance dependence in a nonlinear, non-adapting
system \cite{yu&lee,borst&sompolinsky}. We show this effect in the
analysis of the Abbott-Kepler model neuron.

The identification of a curved or dynamical threshold raises an
issue in reverse correlation analysis regarding the determination of
spike time. For the FitzHugh-Nagumo and Abbott-Kepler models, we
compared two cases: when spike times are identified using a
threshold in voltage, and when the spike times are found using the
crossing of an identified curve that leads to spiking for zero
current input. Surprisingly, the spike triggered averages showed
remarkable differences. For a voltage threshold, the time delay from
the earlier threshold crossing point blurs the estimated filters. It
is shown that using the dynamical threshold leads to a better
fitting of the estimated filters by the system's linear kernels, as
well as improving the sharpness and substantially altering the shape
of the STA.

In all of our discussion here, we have concentrated on the
subthreshold dynamics bringing the system to the point of threshold.
We do not analyze, and our simplified models do not accommodate, the
dynamics of the system immediately after a spike. In the phase plane
picture, spiking reinjects the system into the subthreshold regime
in a nonrandom way, affecting the subsequent probability flux to
threshold, analogous to the one-dimensional reset of the
integrate-and-fire neuron treated in \cite{paninski&lau}. In our
approach, we have assumed that the system has remained below
threshold for long enough that its location in the subthreshold
space has been randomized by the driving current. This corresponds
to an analysis of isolated spikes only, a simplification we have
used before \cite{hh} and employed here for the covariance analysis.
A number of works have treated this issue explicitly with a variety
of methods: treating the interspike interval as the primary symbol
\cite{deruyter&bialek88, spikes}, solving for the interspike
interaction using the interspike intervals \cite{pillow&simoncelli},
simultaneously solving for the linear kernel over stimulus history
and spike history using the autoregressive moving average
\cite{randy&marcARMA,truccolo} and fitting parameters of an explicit
model for the effective post-spike current
\cite{gerstner,keat,paninski&pillow,pillowRetina}. While the
simplification we have used here allows us to find direct
connections between the covariance modes without the confound of the
interspike interaction \cite{hh,i&f}, it is clearly not a complete
model for spiking responses. A first step toward a more complete
spiking model may be to consider the perturbed subthreshold
distributions induced by the influx of trajectories following
spikes. This is in effect a mean field approximation, taking into
account the overall spike rate for a given stimulus ensemble.
Further steps could be taken by introducing a return map
deterministically relating the point of threshold crossing to a
point of return into the subthreshold domain.  The
integrate-and-fire model is the most trivial implementation of such
a return map; an equivalent map in multiple dimensions would
reintroduce all spike trajectories into the subthreshold domain at
an identified point (this is $V=0$ for integrate-and-fire).  Such a
many-to-one map implies that the neuron's state is completely reset
by a spike, which is incorrect for neurons with slow conductances
that modulate spike afterpotentials.  An less degenerate map seems
more appropriate for such cases. Another important step is the
addition of these slow conductances. Using our formalism, such
conductances may be representable simply as additional dimensions of
threshold curvature with corresponding longer-timescale stimulus
filters.

White noise analysis allows the derivation of intuitive functional
models for neural computation: what does the neuron compute? In this
paper we have drawn concrete correspondences between the components
of these functional models and parameters of the underlying
dynamical system.

\appendix

\section{Volterra expansion of a dynamical system}\label{sec:volt-expans-dynam}

Let's consider an $n$-dimensional dynamical system, perturbed by an
external input $I(t)$ as Eq~\eqref{eq:lsystem}. For convenience, we
introduce a dimensionless expansion parameter $\eta$, with which we
will expand the equation. More precisely, Eq~\eqref{eq:lsystem} becomes
\begin{equation}\label{eq:a5}
\dt y_k = f_k ( y_1, y_2, \ldots, y_N ) + \eta\delta_{k0} I( t ),
\end{equation}
We also have the perturbation expansion of $y_k$ as
\begin{equation}\label{eq:a6}
y_k = \xk{0} + \eta \xk{1} + \eta^2 \xk{2} + \cdots.
\end{equation}
By plugging Eq~\eqref{eq:a6} into Eq~\eqref{eq:a5}, we
obtain
\begin{multline*}
\dt
\left(
\eta \xk{1} +
\eta^2 \xk{2} + \cdots \right)
= \eta I(t) \delta_{k0} +
\frac{{\partial} f_k ( y^{( 0 )} ) }{{\partial} y_m} \eta \xx{m}{1}\\
+\frac{1}{2!} \frac{{\partial}^2 f_k ( y^{( 0 )} )}{{\partial} y_m
{\partial} y_n} \eta^2 \xx{m}{1} \xx{n}{1} +
\frac{{\partial} f_k ( y^{( 0 )} )}{{\partial} y_m} {\eta}^2 {\xx{m}{2}}
+ {\cdots} .
\end{multline*}
By comparing two sides order by order, we obtain a series of equations
satisfied by the perturbative expansion at each order as
\[
\cD{km}\xx{m}{1} = I\delta_{k0},\quad
\cD{km} = \delta_{km}\dt -
\left.\frac{\partial f_k ( x^{( 0 )})}{\partial x_m}\right|_{y^{(0)}},
\]
which is Eq~\eqref{eq:7},
\begin{equation}\label{eq:a8}
\cD{km} \xx{m}{2} =  \frac{1}{2!} H_{kmn} \xx{m}{1} \xx{n}{1},\quad
H_{kmn} = \frac{\partial^2 f_k ( y^{( 0 )} )}{\partial y_m \partial y_n},
\end{equation}
which is Eq~\eqref{eq:8}, and so on.

These equations can be solved
recursively by using a kernel $\kep{1} = \cD{}^{-1}$ which therefore
satisfies
\[
\cD{km}\kep{1}_{ml}(t,t') = \delta_{kl}\delta ( t - t' ).
\]
Now the solution of the first order can be written as
\[
\xk{1} ( t ) = \int dt' \, \kep{1}_{k} ( t-t' ) I ( t' ).
\]
where $\kep{1}_{k} = \kep{1}_{k0}$.
Also from Eq~(\ref{eq:a8}),
\begin{eqnarray*}
\xk{2} & = & \frac{1}{2!} \int dt' \, \kep{1}_{kl} ( t-t' )
\left( H_{lmn} \int ds_1 ds_2 \,
\kep{1}_{m} ( t'- s_1 ) \kep{1}_{n} ( t'-s_2 ) I ( s_1 ) I ( s_2 ) \right) \\
& = &
\frac{1}{2!}\int ds_1 ds_2 \, \kep{2}_{k}( t-s_1,t- s_2 ) I ( s_1 ) I ( s_2 ),
\end{eqnarray*}
\[
\kep{2}_{k} ( s_1, s_2 ) =
\int dt'\,
H_{lmn}
\kep{1}_{kl} ( t' )
\kep{1}_{m} ( s_1-t' )
\kep{1}_{n} ( s_2-t' ),
\]
which is Eq~\eqref{eq:2ndkernel}. Since this procedure can be
carried out to higher orders, the higher order kernels are outer
products of the linear kernels.

In addition, the first order kernels are related to each other by
simple time differentiation. To show this, let's rewrite
Eq~(\ref{eq:9}) as follows
\begin{equation}
\kep{1}_{m}(t-t') = \cD{mn}^{-1}\delta_{n0}\delta(t-t').
\end{equation}
Now the derivatives can be computed as
\begin{eqnarray}
\dt \kep{1}_{m} & = & \dt \cD{mn}^{-1}\delta_{n0}\delta(t-t')\nonumber\\
& = & (\dt \delta_{ml} - J_{ml} + J_{ml}) \cD{ln}^{-1}\delta_{n0}\delta(t-t') \nonumber\\
& = & \cD{ml}\cD{ln}^{-1}\delta_{n0}\delta(t-t') + J_{ml}\kep{1}_{l}\nonumber\\
& = & \delta_{m0}\delta(t-t') +  J_{ml}\kep{1}_{l}\label{eq:depsilon}
\end{eqnarray}
This shows that the non-singular part of the first order kernels can
be obtained by the linear combination of the derivatives of other
kernels.

\section{A linear model from the FitzHugh-Nagumo system}\label{sec:exampl-
fitzh-nagumo}
Here we derive the linearization of the FitzHugh-Nagumo model,
beginning with Eqs.~\eqref{eq:fhna} and~\eqref{eq:fhnb}. The fixed
point can be obtained by simultaneously solving the nullcline
equations. We denote it by $(\vo,\wo)$, and we expand the system
around this point.

First, the Jacobian is
\[
J = \left( \begin{array}{cc}
     F' ( \vo ) / \psi & - 1 / \psi\\
     1 & - b
   \end{array} \right),
\]
where $F(V) = V(1-V)(a+V)$, and this defines a linear system
\begin{eqnarray*}
    \dt{V} &  = & \frac{F'(\vo)}{\psi} V - \frac1\psi W + I(t)\\
    \dt{W} & = & V - bW.
\end{eqnarray*}
$J$ has eigenvalues $\lambda_{\pm}$,
\begin{equation}
    \begin{split}
        \lambda_{\pm} & =  \frac{1}{\psi}
        \left[ f_- ( \vo ) \pm \sqrt{f_+ ( \vo )^2 - \psi} \right]\\
         & = -b + \left[ f_+ ( \vo ) \pm \sqrt{f_+ ( \vo )^2 - \psi} \right]\\
         & = -b + \kappa_{\pm}.
    \end{split}
\end{equation}
where $f_{\pm} ( V ) = (F' ( V ) \pm b \psi)/2$. As in
Eq.~\eqref{eq:diagonalize}, $J$ is diagonalized by a matrix $U$,
\[
U = \frac{1}{\lambda_+ - \lambda_-} \left(\begin{array}{cc}
1 & \lambda_- +b\\
-1 & -(\lambda_+ + b)
\end{array}\right).
\]
From Eq.~\eqref{eq:k1final}, we obtain the first order kernels which
solve the linear system
\begin{equation}
    \begin{split}
        \kep{1}_V(t) & = e^{- b t} \dt S (t)H(t),\\
        \kep{1}_W(t) & = e^{- b t} S(t) H(t),\\
    \end{split}
\end{equation}
where $S ( t ) = (e^{\kappa_+ t} - e^{\kappa_- t})/(\kappa_+ -
\kappa_-)$. $\kep{1}_{V,W}(t)$ with our choice of parameters is
drawn in Fig~\ref{fig:FNfirst}a.

\section{Derivation of the Abbott-Kepler model}\label{sec:derivation-abbott-kepler}
In this section, we show the derivation of a two-dimensional neuron model used
in \S\ref{sec:abbott-kepler-model}. For further details, we refer to the
original paper~\cite{abbott&kepler}.

We begin with a Hodgkin-Huxley equation, which is defined by
Eq.~\eqref{eq:ne1} and the following parameters:
\begin{equation}
g_L  = \bar{g}_L,\qquad
g_K  = \bar{g}_K n^4, \qquad
g_{Na} = \bar{g}_{Na} m^3 h.
\end{equation}
\begin{equation}
\tau_{z}(V) \frac{dz}{dt} = \bar{z}(V) - z, \qquad
\tau_{z} = \frac1{\alpha_z+\beta_z}, \qquad
\bar{z} = \frac{\alpha_z}{\alpha_z+\beta_z}, \qquad
z = m, n, h. \label{eq:act-variables}
\end{equation}
\begin{equation}
\begin{split}
 &\alpha_m = \frac{.1(V+40)}{1-\exp[-.1(V+40)]}, \qquad
\beta_m = 4 \exp[−.0556(V + 65)],\\
 &\alpha_h = .07 \exp[−.05(V + 65)], \qquad
\beta_h = \frac{1}{1 + \exp[−.1(V + 35)]},\\
 &\alpha_n = \frac{.01(V+55)}{1-\exp[-.1(V+55)]}, \qquad
\beta_n = .125 \exp[−.0125(V + 65)].
\end{split}
\end{equation}

Now, the key observation is that $\tau_m$ is much smaller than $\tau_{n,h}$
while $\tau_h$ and $\tau_n$ are mutually comparable. Therefore, $m$ can be
approximated by its value at equilibrium, $\bar{m}(V)$, and $h$ and $n$ can be
represented by the same equilibrium voltage, $U$. In other words,
\[
m\approx\bar{m}(V),\qquad h\approx\bar{h}(U),\qquad n\approx\bar{n}(U).
\]
Using this, we obtain Eq.~\eqref{eq:ak1} as
\begin{equation*}
\begin{split}
F &= \sum_{i = L,K,Na} g_i(V-E_i)\\
&\approx \bar{g}_L(V-E_L)
+\bar{g}_K \bar{n}(U)^4(V-E_K)
+\bar{g}_{Na} \bar{m}(V)^3 \bar{h}(U)(V-E_{Na})\\
&= -f(V,U).
\end{split}
\end{equation*}
An equation for $U$, Eq.~\eqref{eq:ak2},
is obtained by requiring that time dependence of
the active current $F$ due to $h$ and $n$ in the Hodgkin-Huxley model is
mimicked by $\bar{h}(U)$ and $\bar{n}(U)$. This implies
\begin{equation}\label{eq:dFdt}
\frac{dF}{dh}\frac{dh}{dt} + \frac{dF}{dn}\frac{dn}{dt}
\approx \left(\frac{\partial f}{\partial\bar{h}} \frac{d\bar{h}}{dU}
+ \frac{\partial f}{\partial\bar{n}} \frac{d\bar{n}}{dU}
\right)\frac{dU}{dt}.
\end{equation}
Again, we approximate Eq.~\eqref{eq:act-variables} in the same way,
\[
    \frac{dz}{dt} \approx
    \frac{1}{\tau_z(V)}\left( \bar{z}(V) - \bar{z}(U)\right), \qquad z = h,n.
\]
Plugging this in Eq.~\eqref{eq:dFdt}, we can solve for $dU/dt$ as a function
of $V$ and $U$, which we denoted by $g(V,U)$ in Eq.~\eqref{eq:ak2} as
\[
g(V,U) = \frac{{\bar{g}_{Na}} (V-E_{Na}) \bar{m}(V)^3 (\bar{h}(V)-\bar{h}(U))
    /\tau_h(V) +4 {\bar{g}_{K}} (V-E_K)
   \bar{n}(U)^3 (\bar{n}(V)-\bar{n}(U))
   /\tau_n(V)}{{\bar{g}_{Na}} (V-E_{Na}) \bar{m}(V)^3 \bar{h}'(U)
   +4 {\bar{g}_{K}} (V-E_K) \bar{n}(U)^3
   \bar{n}'(U)}.
\]


\end{document}